\documentclass[12pt]{article}
\usepackage{latexsym}
\usepackage{amsmath,amsbsy,amssymb}
\usepackage{verbatim}
\usepackage{bm}

\usepackage{curves}
\usepackage{epic}
\usepackage{eepic}
\usepackage{epsfig}

\newcommand{\dd}{\mbox{\rm d}}

\newcommand{\gam}{\gamma}

\newcommand{\dg}{\dagger}

\newcommand{\tl}{\tilde}

\newcommand{\p}{\partial}

\newcommand{\be}{\begin{equation}}
\newcommand{\bear}{\begin{eqnarray}}
\newcommand{\ear}{\end{eqnarray}}
\newcommand{\ee}{\end{equation}}

\newcommand{\lbl}{\label}
\newcommand{\bi}{\bibitem}
\newcommand{\ci}{\cite}

\newcommand{\vs}{\vspace}
\newcommand{\hs}{\hspace}

\newcommand{\bD}{{\bar D}}
\newcommand{\vphi}{\varphi}
\newcommand{\bp}{{\bm p}}
\newcommand{\bk}{{\bm k}}
\newcommand{\bx}{{\bm x}}
\newcommand{\hbp}{{\hat{\bm p}}}
\newcommand{\hbx}{{\hat{\bm x}}}
\newcommand{\vac}{|0 \rangle}
\newcommand{\bmu}{{\bar \mu}}
\newcommand{\bnu}{{\bar \nu}}
\newcommand{\sg}{\sigma}
\newcommand{\ba}{{\bar a}}
\newcommand{\bb}{{\bar b}}

\begin{document}

\begin{center}

\

\vs{.8cm}

\baselineskip .7cm

{\bf \Large A New Approach to the Classical and Quantum Dynamics of Branes } 

\vs{4mm}

\baselineskip .5cm
Matej Pav\v si\v c

Jo\v zef Stefan Institute, Jamova 39,
1000 Ljubljana, Slovenia

e-mail: matej.pavsic@ijs.si

\vs{3mm}

{\bf Abstract}
\end{center}

\baselineskip .45cm

It is shown that the Dirac-nambu-Goto brane can be described as a point particle in
an infinite dimensional brane space with a particular metric. This suggests a generalization
to brane spaces with arbitrary metric, including the ``flat" metric. Then quantization of such
a system is straightforward: it is just like quantization of a bunch of non interacting particles.
This leads us to a system of a continuous set of scalar fields. For a particular choice
of the metric in the space of fields we find that the classical Dirac-Nambu-Goto brane theory
arises as an effective theory of such an underlying quantum field theory. Quantization of
branes is important for the brane world scenarios, and thus for ``quantum gravity".



\baselineskip .55cm

\section{Introduction}

Quantization of the Dirac-Nambu-Goto $p$-brane\,\ci{pBranes}--\ci{DuffBenchmark} is a tough problem that still awaits for a
solution. The problem is rather well understood an resolved in the case of strings
($p=1$)\,\ci{strings}--\ci{strings3}, but not of branes of arbitrary dimensionality $p$. Branes, $p$-branes and
$D$-branes\,\ci{Dbranes,Dbranes2} are important objects in string theory. Since according to the
brane world
scenarios\,\ci{BraneWorld}--\ci{BraneWorld18},\ci{PavsicBook}--\ci{PavsicBrane3} 
a 3-brane sweeping a 4-dimensional world sheet can be
associated with our spacetime, quantization of the brane would as well resolve
the problem of quantum gravity.

In this paper I will show that the Dirac-Nambu-Goto action, governing the dynamics of $p$-branes,
is a special case of an action in an infinite dimensional space ${\cal M}$, for a special choice of the
metric. This suggests that the usual string or $p$-brane theory is embedded in a more general theory=
in which the metric of ${\cal M}$ is arbitrary\,\ci{PavsicBook} in the same sense as is arbitrary the metric of
spacetime in general relativity. In particular, the metric of ${\cal M}$ can be globally diagonal, in which case
we have very special objects that I will call ``{\it flat branes}". A flat brane is like a bunch of
point particles in the absence of any interaction. If we bring an interaction into the game,
then the metric of ${\cal M}$ is no longer globally diagonal, i.e., flat, and the space ${\cal M}$
has non vanishing curvature.

If such a generalized brane is compared with a bunch of interacting particles, then by
analogy, the generic metric of the particle configuration space ${\cal C}$ should be like the
generic metric of the brane space ${\cal M}$. Such a reasoning suggests that the usual
many particle interacting theory and its quantization should be generalized so that
in the limit of continuous bunch of particles it would match  the theory of general branes
and the curved brane space ${\cal M}$. In the case of flat brane space ${\cal M}$ and
flat particle configuration space ${\cal C}$, the generalized brane theory that allows
for the diagonal metric of ${\cal M}$, matches the usual theory of non interacting point
particles, where  the metric of ${\cal C}$ is also diagonal.

Quantized theory of a generalized brane should thus start just as the quantization
of many particle systems: in flat brane space ${\cal M}$. By analogy, quantized theory
of an interacting many particle system should be formulated not in 4-dimensional
spacetime, but in the many dimensional configuration space ${\cal C}$. The corresponding
quantum fields are then functions of position in ${\cal C}$. As a model we consider
the theory of a scalar field in a multidimensional configuration space. The action for
such a system can be reduced to the action for a system of scalar fields describing
non interacting distinguishable particles. In the continuum limit we have a bunch
of non interacting particles forming a flat brane. Such a bunch
of particles---a flat brane---is described by a continuous set of scalar fields, differing
in the brane parameters $\sg^\ba$, $\ba=1,2,...,p$, the metric $s(\sg,\sg')$ of the
field space being diagonal ($\delta$-function like). If we replace the diagonal
metric with a more general one, then we have interactions amongst the
brane constituent ``particles".

We have found a particular metric
$s(\sg,\sg')= (1 + \lambda \p^\ba \p_\ba) \delta^p (\sg - \sg')$ which enables
straightforward exact calculations and quantization. The inverse of such
a metric is the propagator in the space $\lbrace \sg^\ba \rbrace$. The theory becomes
a theory of the scalar field $\vphi(\sg^\ba,x^\mu)$ whose arguments are
not only spacetime coordinates $x^\mu$, but also the brane coordinates
$\sg^\ba$. It is straightforward to compute the exact Hamiltonian and
momentum operator ${\hat \bp}$. We then calculated how the expectation
value of $\hbp$ in a state which is the product of single particle wave packet
profiles changes with time. We obtained the equations of motion for the
wave packet centroid coordinates ${\bar X}_\bmu (\sg)$ that match
the classical brane equations of motion in the case when the determinant
 $(-{\bar \gam})$ of the brane's induced metric $\gam_{\ba \bb}$ is equal to $1$. If we generalize
 the field space metric according to
 $s(\sg,\sg')= \left (1 + \lambda \p^\ba (\sqrt{-{\bar \gam}} \gam^{\ba \bb} \p_\bb)
 \right ) \delta^p (\sg - \sg')$, then as the expectation value we obtain
 exactly the classical brane equations of motion. The classical brane theory is thus
 obtained as an effective theory of our underlying quantum field theory of
 a continuous system of scalar fields for a particular choice of the field
 space metric $s(\sg,\sg')$. For a different choice of $s(\sg,\sg')$ we would
 obtain a different effective classical brane---in agreement with our starting
 assumption that the brane space of a classical brane can have in principle
 an arbitrary metric, not necessarily the metric that gives the Dirac-Nambu-Goto
 brane.

\section{A brane as a ``point particle" in an infinite dimensional space}

The Dirac-Nambu-Goto action for a $p$-brane is
\be
  I = \kappa \int \dd^{p+1} \xi \, (-\gam)^{1/2},
\lbl{2.1}
\ee
Here $\gam \equiv {\rm det} \,\gam_{ab}$, $\gam_{ab} \equiv \p_a X^\mu \p_b X_\mu$,
where $X^\mu (\xi^a)$, $\mu=0,1,2,...,D-1$, $a=0,1,2,...,p$, are the embedding
functions of the world volume swept by the $p$-brane, and $\kappa$ is the
brane tension.

A action that is equivalent to (\ref{2.1}) is the Schild action\,\ci{Schild}
\be
  I_{\rm Schild} =\frac{\kappa}{2 k} \int \dd^{p+1}\xi \, (-\gam),
\lbl{2.1a}
\ee
in which the determinant of the induced metric occurs without the square root.
This is a gauge fixed action. The equations of motion give $\p_a (-\gam) =0$,
which means that $(-\gam) =C$, where $C$ is a constant. If we choose a gauge so
that
\be
  (-\gam) = C = k^2
\lbl{2.1b}
\ee
then the momentum ${{\pi}_\mu}^c = \kappa \sqrt{-\gam} \, \p^c X_\mu$ derived from
the Dirac-Nambu-Goto action (\ref{2.1}) is equal to the momentum
$\kappa (-\gam) \p^c X_\mu/k$ derived from the Schild action. This is the reason
why in (\ref{2.1a}) we included an additional constant factor $1/(2k)$. 

The action (\ref{2.1}) is invariant under reparametrizations of the parameters
$\xi^a \equiv (\tau, \sigma^{\bar a})$, ${\bar a}=1,2,...,p$. We can choose a particular
gauge (choice of $\xi^a$) in which the determinant factorizes according to
\be
   (-\gam) = {\dot X}^2 (-{\bar \gam}),
\lbl{2.2}
\ee
where ${\dot X}^2 \equiv {\dot X}^\mu {\dot X}_\mu$, ${\dot X}^\mu \equiv \p X^\mu/\p \tau$,
and ${\bar \gam} \equiv {\rm det}\, \p_{\bar a}  X^\mu \p_{\bar b} X_\mu$, ${\bar a},{\bar b} = 1,2,...,p$.
The action (\ref{2.1}) then becomes
\be
   I = \kappa \int \dd \tau \, \dd^p \sigma \, \sqrt{{\dot X}^2} \sqrt{-{\bar \gam}}.
\lbl{2.3}
\ee

The Schild action (\ref{2.1a})
is invariant under those coordinate transformations of $\xi^a$ which preserve the determinant $\gam$.
Also for the Schild action we can choose a gauge in which holds the factorization (\ref{2.2}).
Then Eq.\,(\ref{2.1a}) becomes
\be
  I_{\rm Schild} =\frac{\kappa}{2 k} \int \dd \tau \, \dd^p \sigma \, {\dot X}^2 (-{\bar \gam})
\lbl{2.1c}
\ee
This can be written as
\be
  I = \frac{\kappa}{2 k} \int \dd \tau \, \dd^p \sigma \,\dd^p \sigma' \, (-{\bar \gam})\,
  \eta_{\mu \nu} \, \delta(\sigma-\sigma')\,{\dot X}^\mu (\tau,\sigma){\dot X}^\nu (\tau,\sigma').
\lbl{2.4}
\ee
At every $\tau$, the integrand is a quadratic form in an infinite dimensional space
with the metric
\be
  \rho_{\mu \nu} (\sigma,\sigma') = (-{\bar \gam})\, \eta_{\mu \nu} \, \delta(\sigma-\sigma').
\lbl{2.5}
\ee
Introducing the compact notation
\be
   {\dot X}^{\mu (\sigma)}(\tau) \equiv {\dot X}^\mu (\tau,\sigma)\, , ~~~~~~
   \rho_{\mu(\sigma)\nu(\sigma)} \equiv \rho_{\mu \nu} (\sigma,\sigma'),
   \lbl{2.6}
\ee
the action (\ref{2.4}) reads\footnote{
We use the generalization of Einstein's summation convention, so that not only
summation over the repeated indices $\mu,\nu$, but also the integration over the
repeated continuous indices $(\sigma)$, $(\sigma')$ is assumed.}
\be
   I _{\rm Schild}= \frac{\kappa}{2 {k}} \int \dd \tau \, \rho_{\mu(\sigma)\nu(\sigma')}  {\dot X}^{\mu (\sigma)}(\tau)
  {\dot X}^{\nu (\sigma')}(\tau).
\lbl{2.7}
\ee
 
 The momentum derived from the action (\ref{2.1c}) is
\be
  {p}_\mu (\sigma) = \frac{\kappa (-{\bar \gam}) {\dot X}_\mu} {k} = \frac{\kappa \sqrt{-{\bar \gam}}  {\dot X}_\mu} { \sqrt{{\dot X}^\mu {\dot X}_\mu}},
 \lbl{2.7b}
\ee
where we have taken into account
\be
  k = \sqrt{{\dot X}^2} \sqrt{-{\bar \gam}} ,
\lbl{2.7bb}
\ee
which follows from (\ref{2.1b}) and (\ref{2.2}).

For the momentum belonging to the action (\ref{2.7})  we obtain the expression which equals to (\ref{2.7b}):
\be
  {p}_{\mu(\sigma)} =\frac{\kappa}{k} \, \rho_{\mu(\sigma)\nu(\sigma')} {\dot X}^{\nu(\sigma')}
  = \frac{\kappa}{k} {\dot X}_{\mu (\sigma)} = \frac{\kappa}{k} \,(-{\bar \gam}) {\dot X}_\mu (\sigma),
\lbl{2.7c}
\ee
where ${\dot X}_{\mu (\sigma)} = \rho_{\mu(\sigma)\nu(\sigma')} {\dot X}^{\nu(\sigma')}$,
and ${\dot X}_\mu (\sigma) = \eta_{\mu \nu} {\dot X}^\nu$, ${\dot X}^\mu (\sigma) 
\equiv {\dot X}^{\mu (\sigma)}$. We identify $p_{\mu(\sigma)} \equiv p_\mu (\sigma)$, where
\be
  p^{\mu (\sigma)} = \rho^{\mu(\sigma) \nu(\sigma')} p_{\nu(\sigma')} = \frac{p^\mu (\sigma)}{(-{\bar \gam})},
\lbl{I1}
\ee
where $p^\mu (\sigma) = \eta^{\mu \nu} p_\nu (\sigma)$.

From the definition (\ref{2.7b}) of the momentum we obtain the following constraint:
\be
  p_\mu (\sigma) p^\mu (\sigma) = \eta^{\mu \nu} p_\mu (\sg) p_\nu (\sg) = \kappa^2 (-{\bar \gam}).
\lbl{I2}
\ee
We also have
\be
  p_{\mu(\sigma)} p^{\mu(\sigma)} = \rho^{\mu(\sg) \nu(\sg')} p_{\mu(\sg)} p_{\nu(\sg')} =  {\tl \kappa}^2,
\lbl{I3}
\ee
where
\be
  {\tl \kappa}^2 = \int \kappa^2\, \dd \sigma .
\lbl{I3a}
\ee
In deriving Eq.\,(\ref{I3}) we wrote $p_{\mu(\sigma)} p^{\mu(\sigma)} = \rho^{\mu(\sigma) \nu(\sigma')} 
p_{\mu(\sigma)} p_{\nu(\sigma')}$, where $\rho^{\mu(\sigma) \nu(\sigma')} = \eta^{\mu \nu} \delta(\sigma-\sigma')/ (-{\bar \gam})$,
and used (\ref{I2}).

Eq.\,(\ref{2.7}) can then be written in the form
\be
   I _{\rm Schild}= \frac{\tl \kappa}{2 {\tl k}} \int \dd \tau \, \rho_{\mu(\sigma)\nu(\sigma')}  {\dot X}^{\mu (\sigma)}(\tau)
  {\dot X}^{\nu (\sigma')}(\tau),
\lbl{I4}
\ee 
where
\be
  {\tl k}^2 = \rho_{\mu(\sigma) \nu(\sigma')} {\dot X}^{\mu(\sigma)} {\dot X}^{\nu(\sigma')} =
  \int \dd \sigma (-{\bar \gam}) {\dot X}^\mu {\dot X}^\nu = \int k^2 \dd \sigma.
\lbl{I5}
\ee
Eq.\, (\ref{I4}) is a generalization to infinite dimensions of the Schild action
\be
   I_{\rm Schild} = \frac{m}{2 k} \int \dd \tau \, g_{\mu \nu} {\dot X}^\mu {\dot X}^\nu 
\lbl{2.8}
\ee
for a relativistic point particle in a curved spacetime with the metric $g_{\mu \nu}$.
The latter action is a gauge fixed action for a relativistic point particle, which
is described by the reparametrization invariant action
\be
  I = m \int \dd \tau \, (g_{\mu\nu} {\dot X}^\mu {\dot X}^\nu )^{1/2} .
\lbl{2.8a}
\ee
Analogously, instead of (\ref{I4}), we can take the action
\be
   I = {\tl \kappa} \int \dd \tau  \left ( \rho_{\mu(\sigma)\nu(\sigma)}  {\dot X}^{\mu (\sigma)}(\tau)
  {\dot X}^{\nu (\sigma')}(\tau) \right )^{1/2}.
\lbl{2.8b}
\ee

From (\ref{I3a}) and (\ref{I5}) it follows that ${\tl \kappa}/{\tl k} = k/\kappa$, i.e.,
\be
  \frac{\tl \kappa}{\sqrt{{\dot X}^{\mu(\sigma)} {\dot X}_{\mu(\sigma)}}} = 
  \frac{\kappa}{\sqrt{(-{\bar \gam}) {\dot X}^\mu {\dot X}_\mu}}.
\lbl{I6}
\ee
This relation is valid not only in the gauge in which $\sqrt{{\dot X}^2} \sqrt{-{\bar \gam}} = k = constant$,
but also in an arbitrary gauge. The same is true for the constraints (\ref{I2}) and (\ref{I3}).

Similarly to a point particle being described  at every $\tau$ by a finite set of coordinates
$x^\mu$ of a point (event) in a finite dimensional spacetime, a brane is described at every $\tau$
by an infinite set of coordinates $x^{\mu(\sigma)} \equiv x^\mu (\sigma)$ of a point in
an infinite dimensional space, the so called brane space ${\cal M}$\,\ci{PavsicBook,BraneSpace}.
As $\tau$ monotonically increases, the brane traces a worldline in ${\cal M}$,
described by the parametric equation $x^{\mu(\sigma)} = X^{\mu(\sigma)} (\tau)$,
which is a mapping from the 1-dimensional space of the parameter $\tau$ into
an infinite dimensional brane space ${\cal M}$ whose points are denoted by an
infinite set of coordinates $x^{\mu (\sigma)}$. The latter coordinates describe
a {\it brane event} in an analogous way as the coordinates $x^\mu$, $\mu=0,1,2,3$
described a {\it point particle event} in spacetime. The distinction between the lower case
$x^{\mu (\sigma)}$ and the capital $X^{\mu (\sigma)}$ is the distinction between
coordinates of the brane space ${\cal M}$ and $\tau$-dependent functions
$X^{\mu (\sigma)} \equiv X^{\mu(\sigma)} (\tau)$. The derivative
${\dot X}^{\mu(\sigma)}(\tau)$ is the velocity in ${\cal M}$.

Here $x^{\mu(\sigma)} \equiv x^\mu (\sigma)$ denotes a {\it kinematically possible brane}.
We postulate that distinct functions $x^\mu (\sigma)$ can describe physically different
branes, even in the case in which they are related to each other by a diffeomorphism
$\sigma^i \rightarrow \sigma'^{\bar a} = f^{\bar a} (\sigma)$. In such a case, a diffeomorphism is
{\it active}: it transforms one brane into another brane that is physically different in
the sense that its points are tangentially displaced while the mathematical surface of
both branes are the same\footnote{
See more detailed explanation and figures of Refs.\,\ci{PavsicBook,BraneSpace}}.
On the other hand, a {\it passive diffeomorhism} only relables the parameters $\sigma^{\bar a}$
into new parameters $\sigma'^{\bar a}$, whereas the brane remains the same.

Consideration of the branes related by active diffeomorphisms as distinct kinematically
possible objects is a crucial step that enables a formulation of the generalized brane
theory. Then the tensor calculus of general relativity can be straightforwardly generalized
to infinite dimensions. A point particle event with coordinates $x^\mu$ is analogous to a brane event 
with coordinates $x^{\mu(\sigma)}$. A diffeomorphism in spacetime,
\be
  x^\mu \rightarrow x'^\mu = f^\mu (x^\nu),
\lbl{2.9}
\ee
is analogous to a diffeomorphism
\be
  x^{\mu(\sigma)} \rightarrow x'^{\mu(\sigma)}= F^{\mu(\sigma)} (x^{\nu(\sigma)}),
\lbl{2.10}
\ee
i.e.,
\be
  x^\mu (\sigma) \rightarrow x'^\mu (\sigma) = F^\mu (\sigma) [x^\nu (\sigma)],
\lbl{2.11}
\ee
where $F^\mu (\sigma) [x^\nu (\sigma)]$ are functionals of $x^\nu (\sigma)$.
The new ${\cal M}$-space coordinates $x'^{\mu(\sigma)}$ are functions of
the old ${\cal M}$-space coordinates  $x^{\mu(\sigma)}$, i.e., the embedding
functions $x'^\mu (\sigma)$ are functionals of the old embedding functions
$x^\mu (\sigma)$. Both those diffeomorphisms can be either {\it passive}
or {\it active}. If interpreted passively, then Eq.\,(\ref{2.10}) means that
the same brane is described either by ${\cal M}$-space coordinates\footnote{
Notice that the prime does not mean a derivative, but new quantities.}
$x^{\mu(\sigma)}$ or $x'^{\mu(\sigma)}$.

Diffeomorphisms in the brane space ${\cal M}$ include the diffeomorphisms
within the brane as well:
\bear
  &&\sigma^{\bar a} \rightarrow \sigma'^{\bar a} = f^ {\bar a} (\sigma),\lbl{2.12}\\
  \Rightarrow &&x^\mu (\sigma) \rightarrow x^\mu (f(\sigma)) = x'^\mu (\sigma').
\lbl{2.13}
\ear
In the latter expression we can rename $\sigma'$ into $\sigma$ and write
$x'^\mu (\sigma)$ instead of $X'^\mu (\sigma')$. Eq.\,(\ref{2.13}) means
that the brane space coordinates $x^{\mu(\sigma)}$ transform into new
brane space coordinates $x'^{\mu(\sigma)}$. Those new coordinates can be
interpreted in the {\it passive sense}, namely that they describe the same brane,
or in the {\it active sense}, namely that  they describe a different, i.e., tangentially
deformed, brane.

Tensor calculus in brane space ${\cal M}$ is analogous to that in spacetime.
For instance, under a diffeomorphism (\ref{2.10}) the velocity and the metric
transform as
\be
  {\dot X}'^{\mu(\sigma)} = \frac{\p x'^{\mu (\sigma)}}{\p x^{\nu (\sigma')}} 
  {\dot X}^{\nu (\sigma')},
\lbl{2.14}
\ee
\be
  \rho'_{\mu(\sigma)\nu(\sigma')} = \frac{\p x^{\alpha(\sigma'')}}{\p x^{\mu(\sigma)}}
  \frac{\p x^{\beta(\sigma''')}}{\p x^{\nu(\sigma')}}  \rho_{\alpha(\sigma'')\beta(\sigma''')}.
\lbl{2.15}
\ee
Here we use the following notation for functional derivatives:
\be
  \p_{\mu(\sigma)} \equiv \frac{\p}{\p x^{\mu(\sigma)}} \equiv \frac{\delta}{\delta x^\mu (\sigma)}.
\lbl{2.16}
\ee

In general, the metric of ${\cal M}$ need not be of the form (\ref{2.5}). Moreover, it can
be a metric that is not equivalent to (\ref{2.5}) via a diffeomorphism in ${\cal M}$;
it can be a completely different metric. In Refs.\,\ci{PavsicBook,BraneSpace}
it was proposed that the metric of ${\cal M}$ is dynamical like the metric of spacetime
in general relativity.

The equations of motion derived from (\ref{2.8b}) are
\be
  \frac{\p I}{\p X^{\mu(\sigma)}}= {\tl \kappa} \frac{\dd}{\dd \tau}
  \left ( \frac{{\dot X}_{\mu(\sigma)}}{\sqrt{{\dot {\tl X}}^2}} \right ) -\frac{{\tl \kappa}}{2}\,
  \p_{\mu (\sigma)}  \rho_{\alpha(\sigma')\beta(\sigma'')}\,
  \frac{{\dot X}^{\alpha(\sigma')} {\dot X}^{\beta(\sigma'')}}{\sqrt{{\dot {\tl X}}^2}} = 0,
\lbl{2.16a}
\ee
where ${\dot {\tl X}}^2 \equiv {\dot X}^{\mu(\sigma)} {\dot X}_{\mu(\sigma)} =
\rho_{\mu(\sigma) \nu(\sigma')} {\dot X}^{\mu(\sigma)} {\dot X}^{\nu(\sigma')}$. This
is the geodesic equation in the brane space ${\cal M}$, and, after using
${\dot X}_{\mu(\sigma)}=\rho_{\mu(\sigma) \nu (\sigma)} {\dot X}^{\nu (\sigma)}$
it can be written in the form
\be
  \frac{1}{\sqrt{{\dot {\tl X}}^2}}  \frac{\dd}{\dd \tau}
  \left ( \frac{{\dot X}^{\mu(\sigma)}}{\sqrt{{\dot {\tl X}}^2}} \right )
  + \frac{\Gamma_{\alpha(\sigma') \beta(\sigma'')}^{\mu(\sigma)}
   {\dot X}^{\alpha(\sigma')} {\dot X}^{\beta(\sigma'')}}{{\dot {\tl X}}^2} = 0,
\lbl{2.16b}
\ee
where
\be
   \Gamma_{\alpha(\sigma') \beta(\sigma'')}^{\,\mu(\sg)}
   = \frac{1}{2}\rho^{\mu(\sigma) \gamma(\sigma''')}
   (\rho_{\gamma(\sigma''') \alpha(\sigma'),\beta(\sigma'')} + \rho_{\gamma(\sigma''') \beta(\sigma''),
   \alpha(\sigma')} - \rho_{\alpha(\sigma') \beta(\sigma''),\gamma(\sigma''')} )
\lbl{2.16c}
\ee
is the connection in ${\cal M}$, with comma denoting the functional derivative. The inverse
metric $\rho^{\mu(\sigma) \nu(\sigma')}$ is given
by $\rho^{\mu(\sigma) \nu(\sigma')}\rho_{\nu (\sigma') \alpha(\sigma'')} 
= {\delta^{\nu(\sigma')}}_{\alpha(\sigma'')}$. In the usual notation this read
$\int \dd^p \sigma' \, \rho_{\rm inv}^{\mu \nu} (\sigma,\sigma') \rho_{\mu \nu} (\sigma,\sigma')
= {\delta^\nu}_\mu \delta (\sigma'-\sigma'')$.

Eq.\,(\ref{2.16a}) holds for any metric. For the particular metric (\ref{2.5}), the action (\ref{2.8b})
reads
\be
  I= {\tl \kappa} \int \dd \tau \left ( \int \dd^p \sigma\, (-{\bar \gam}) {\dot X}^2 \right )^{1/2}.
\lbl{2.16d}
\ee
This can be written as $I= {\tl \kappa} \int \dd \tau \, {\cal L}[{\dot X}^\mu (\sigma), X^\mu (\sigma)]$,
where the Lagrangian
\be
  {\cal L}[{\dot X}^\mu (\sigma), X^\mu (\sigma)]  = \left ( \int \dd^p \sigma\, (-{\bar \gam}) {\dot X}^2 \right )^{1/2}
\lbl{2.16e}
\ee
is a functional of infinite dimensional velocities and coordinates.

The Euler-Lagrange equations
\be
   \frac{\dd}{\dd \tau} \frac{\delta {\cal L}}{\delta {\dot X}^\mu (\sigma)} - \frac{\delta {\cal L}}{\delta X^\mu (\sigma)} =0
\lbl{2.16f}
\ee
give
\be
   \frac{\dd }{\dd \tau} \left ( \frac{\tl \kappa}{\sqrt{{\dot {\tl X}}^2}} (-{\bar \gam}) {\dot X}_\mu \right )
  + \p_{\bar a} \left ( \frac{{\tl \kappa} (-{\bar \gam}) {\dot X}^2 \p^{\bar a} X_\mu}{\sqrt{{\dot {\tl X}}^2}} \right ) = 0,
\lbl{2.16g}
\ee
where ${\dot {\tl X}}^2 \equiv {\dot X}^{\mu(\sigma)} {\dot X}_{\mu(\sigma)} =\int \dd^p \sigma\, (-{\bar \gam}) {\dot X}^2$.

Using (\ref{I6}), the equation of motion (\ref{2.16g}) becomes
\be
   \frac{\dd }{\dd \tau} \left ( \frac{\kappa \sqrt{-{\bar \gam}}}{\sqrt{{\dot X}^2}} {\dot X}_\mu \right )
  + \p_{\bar a} \left ( \kappa \sqrt{-{\bar \gam}} \sqrt{{\dot X}^2} \p^{\bar a} X_\mu \right ) = 0,
\lbl{2.16ge}
\ee
This is the same equation as that derived from the Dirac-Nambu-Goto action (\ref{2.3}). We have thus
verified that the action (\ref{2.16d}), which is just (\ref{2.8b}) for a particular metric (\ref{2.5}),
gives the same equations of motion
as the action (\ref{2.3}). The form of the action (\ref{2.8b}) suggests that in general the metric can be arbitrary,
either a ``curved" or a ``flat" metric, including the brane space analog  of the metric $\eta_{\mu \nu}$

\section{Flat brane space: a brane as a bunch of non interacting point particles}

The reasoning at the end of the last section suggests that we should start formulating the brane
theory with the most simple metric, i.e.,
\be
   \rho_{\mu(\sigma) \nu(\sigma')} = \eta_{\mu(\sigma) \nu(\sigma')} 
   =\eta_ {\mu \nu} \delta (\sigma-\sigma'),
\lbl{2.17}
\ee
which is the metric of {\it flat} brane space ${\cal M}$. With such a metric
the action (\ref{2.8b}) becomes
\be
  I = {\tl \kappa} \int \dd \tau \, \left ( \int \dd^p \, \sigma \, \eta_{\mu \nu} 
    {\dot X}^\mu (\tau,\sigma) {\dot X}^\nu (\tau,\sigma) \right )^{1/2} .
\lbl{2.18}
\ee
This is an action for a brane in flat background space ${\cal M}$. Such a brane we will call {\it flat brane}.
The action (\ref{2.18}) is not invariant under general coordinate transformations 
(\ref{2.10}) in ${\cal M}$-space.
Under a diffeomorphism (\ref{2.10}) the metric (\ref{2.17}) occurring
in the latter action transforms according to (\ref{2.15}) into a new metric.
A diffeomorphism (\ref{2.13}) is just a particular diffeomorphism in ${\cal M}$-space,
and the action (\ref{2.8b}) is invariant under (\ref{2.13}), and of the same form.
This shows that we need not to worry that (\ref{2.18}) does not
contain a square root of the determinant of the metric in the $\sigma^{\bar a}$ space,
because the action (\ref{2.18}) is a particular case of the  action (\ref{2.8b}) in which
the metric is fixed according to (\ref{2.17}), and which
is invariant and covariant from the ${\cal M}$-space point of
view.

The equations of motion derived from (\ref{2.18}) are
\be
  \frac{\dd}{\dd \tau} \left ( \frac{{\dot X}^\mu (\tau,\sigma)}{{({\tl {\dot X}}^2)}^{1/2}}
  \right ) = 0,
\lbl{2.19}
\ee
where now we have ${\tl {\dot X}}^2 \equiv   {\dot X}^{\nu(\sigma)} {\dot X}_{\nu(\sigma)} =
\int \dd^p \sigma \, {\dot X}^\mu (\sigma) {\dot X}^\nu (\sigma) \eta_{\mu \nu}$. In a
gauge in which ${\tl {\dot X}}^2 = 1$, the equations of motion read
\be
   {\ddot X}^\mu (\tau,\sigma) = 0,
\lbl{2.19a}
\ee
and their solution is
\be
  X^\mu (\tau,\sigma) = v^\mu (\sigma) \tau + X_0^\mu (\sigma).
\lbl{2.20}
\ee
This is a bunch of straight worldlines. In other words, Eq.\,(\ref{2.20})
represents a continuum limit of a system of non-interacting point
particles, tracing straight worldlines.

Quantization of the system described by  the action (\ref{2.18}) can be performed
in analogous way as the quantization of the point particle in flat spacetime.
Eq.\,(\ref{2.18}) implies the constraint
\be
 \hs{1.5cm} p^{\mu(\sigma)} p_{\mu (\sigma)} - {\tl \kappa}^2 = 0~,
  ~~~~~~~p_{\mu(\sigma)} = \frac{{\tl \kappa} {\dot X}_{\mu(\sigma)}}{\sqrt{{\tl {\dot X}}^2}},
\lbl{2.23}
\ee
where
\be
   p^{\mu(\sigma)} p_{\mu(\sigma)} = \rho_{\mu(\sigma)\nu(\sigma')} p^{\mu(\sigma)} p^{\nu(\sigma')}
   = \int \dd^p \sigma\, \eta_{\mu \nu} p^\mu (\sigma) p^\nu (\sigma),
\lbl{2.25}
\ee
Upon quantization, Eq.\,(\ref{2.23}) becomes the generalized Klein-Gordon equation,
\be
   ({\hat p}^{\mu(\sigma)} {\hat p}_{\mu(\sigma)} - {\tl \kappa}^2 ) \phi (x^{\nu(\sigma)}) = 0~,
\lbl{2.24}
\ee
\be
      ~~~~~~~{\hat p}_{\mu(\sigma)} \equiv -i \p_{\mu(\sigma)}
   \equiv -i \frac{\p}{\p x^{\mu(\sigma)}}= -i \frac{\delta}{\delta x^\mu (\sigma)},
\lbl{2.23a}
\ee
in which the field
\be
  \phi(x^{\nu(\sigma)}) \equiv \phi[x^\mu (\sigma)]
\lbl{2.26}
\ee
is a functional of the brane's embedding functions $x^\mu (\sigma)$.

The corresponding action for the  equation (\ref{2.24}) is
\be
  I[\phi(x^{\mu(\sigma)}] = \frac{1}{2} \int {\cal D} x^{\nu(\sigma)} 
  (\p_{\mu(\sigma)} \phi \, \p^{\mu(\sigma)} \phi - {\tl \kappa}^2 \phi^2).
\lbl{2.28}
\ee
Explicitly, Eq.\,(\ref{2.24}) reads
\be
   \left ( \p_{\mu(\sigma)} \p^{\mu (\sigma)} + {\tl \kappa}^2 \right ) \phi = 0,
\lbl{2.29}
\ee
which in the usual notation reads
\be
\left ( \int \dd^p \sigma \dd^p \sigma' \, \eta^{\mu \nu} \delta (\sigma-\sigma') \,
  \frac{\delta^2}{\delta x^\mu (\sigma) \delta x^\nu (\sigma')} + {\tl \kappa}^2
 \right ) \phi = 0.
\lbl{2.30}
\ee

A particular solution is
\be
  \phi = {\rm e}^{i p_{\mu(\sigma)} x^{\mu (\sigma)}} ,
\lbl{2.31}
\ee
where the momentum eigenvalue $p_{\mu(\sigma)}$ satisfies the constraint
(\ref{2.23}).

A general solution of Eq.\,(\ref{2.29}) is
\be
  \phi(x^{\mu(\sigma)}) = \int {\cal D} p \, c(p) \, {\rm e}^{i p_{\mu(\sigma)} x^{\mu (\sigma)}}
  \delta (p_{\mu(\sigma)} p^{\mu(\sigma)} - {\tl \kappa}^2 ).
\lbl{2.32}
\ee
 
Upon second quantization,  $\phi(x^{\mu(\sigma)})$ becomes the operator
that creates or annihilates a brane with coordinates $x^{\mu(\sigma)} \equiv x^\mu (\sigma)$.
Because we consider the flat brane space with the metric (\ref{2.17}), a brane with
coordinates $x^{\mu (\sigma)}$ is in fact a bunch of non interacting point particles,
i.e., a continuous limit of a many particle system. 

The coefficients $c(p) \equiv c(p_{\mu(\sigma)})$ determine the profile of the wave
packet. Let us consider the case in which
\be
  c(p_{\mu(\sigma)}) = c(p_{\mu(\sigma)})|_{\sigma \in (0,\Delta \sigma)}
  \delta (p_{\mu(\sigma)} - p_{0\mu(\sigma)})|_{\sigma \in (\Delta \sigma,L)}
\lbl{3.1}
\ee
This means that the brane momentum in the interval from $\sigma=0$ to
$\sigma = \Delta \sigma$ is undetermined, whereas in the interval $\sigma \in
(\Delta \sigma,L)$ is sharply determined, so that it is equal to $p_{0 \mu(\sigma)}$.

If we insert (\ref{3.1}) into the general solution (\ref{2.32}), we obtain
\bear
  &&\phi(x^{\mu (\sigma)}) = A \int \prod_{\mu,\sigma \in (0,\Delta \sigma)} \dd p_{\mu(\sigma)}
  \,{\rm exp} \left [ i \int_0^{\Delta \sigma} p_\mu (\sigma) x^\mu (\sigma) \dd \sigma \right]\nonumber\\
  &&\hs{2cm}c(p_\mu (\sigma))|_{\sigma\in (0,\Delta \sigma)}
  \delta \left[ \int_0^{\Delta \sigma} (p_\mu (\sigma) p^\mu (\sigma) - \kappa^2) \dd \sigma \right ],
\lbl{3.2}
\ear
where
\be
  A = {\rm exp} \left [ i \int_{\Delta \sigma}^L p_{0 \mu} (\sigma) x^\mu (\sigma) \dd \sigma \right]
\lbl{3.2a}
\ee
is a phase factor. In arriving at (\ref{3.2}) we have used
$$
  \int {\cal D} p_{\mu(\sigma)}|_{\sigma \in (\Delta \sigma,L)} 
  \delta (p_{\mu(\sigma)} p^{\mu(\sigma)} - {\tl \kappa}^2)
  \delta (p_{\mu(\sigma)} - p_{0 \mu(\sigma)} )|_{\sigma \in (\Delta \sigma,L)} $$
\be
   = \delta \left [ \int_0^{\Delta \sigma} p_\mu (\sigma) p^\mu (\sigma)  \dd \sigma+
    \int_{\Delta \sigma}^L p_{0 \mu} (\sigma) p_0^\mu (\sigma) \dd \sigma  - {\tl \kappa}^2 \right ]
   = \delta \left ( \int_0^{\Delta \sigma} p_\mu (\sigma) p^\mu (\sigma) - \kappa^2 \sigma \right ) \dd \sigma,
\lbl{3.3}
\ee
where
\be
   \int_0^{\Delta \sigma} \kappa^2 \dd \sigma = {\tl \kappa}^2 
   - \int_{\Delta \sigma}^L p_{0 \mu} (\sigma) p_0^\mu (\sigma) \dd \sigma  .
\lbl{3.4}
\ee
This is consistent with ${\tl \kappa}^2 = \int_0^L \kappa^2 \dd \sigma$
(see Eq.\,(\ref{I3a}).

Eq.\,(\ref{3.2}) is a solution of the generalized Klein-Gordon equation
(\ref{2.24}). But because the momenta $p_{\mu(\sigma)} \equiv p_\mu (\sigma)$
for $\sigma \in (\Delta \sigma,L)$ have been integrated out, the field (\ref{3.2}) is a solution of the
Klein-Gordon equation restricted to $\sigma \in (0,\Delta \sigma)$ as well:
\be
\left [\int_0^{\Delta \sigma} ({\hat p}_\mu (\sigma) {\hat p}^\mu (\sigma) - \kappa^2) \dd \sigma \right ]
\phi(x^{\mu(\sigma)} ) = 0.
\lbl{3.5}
\ee

The $\delta$-function constraint in Eq.\,(\ref{3.2}) can be written as
\be 
  \int_0^{\Delta \sigma} (p_\mu (\sigma) p^\mu (\sigma) - \kappa^2) \dd \sigma 
  \approx  (p_\mu (\sigma) p^\mu (\sigma) - \kappa^2) \Delta \sigma = 0.
\lbl{3.6}
\ee
Multiplying the latter expression by $\Delta \sigma$ and introducing
$p_\mu = p_\mu (\sigma) \Delta \sigma$, $\kappa \Delta \sigma = m$, we
obtain
\be
  p_\mu p^\mu - m^2 = 0,
\lbl{3.7}
\ee
which is the constraint among the point particle momenta.

Not only in the generalized, but also in the restricted Klein-Gordon equation (\ref{3.5}),
the momentum operator is the functional derivative (\ref{2.23a}). By proceeding
in the analogous way as in Eqs.\,(\ref{3.6}),(\ref{3.7}), we obtain
\be
  ({\hat p}_\mu {\hat p}^\mu - m^2 ) \varphi (x^\mu) = 0,
\lbl{3.8}
\ee
where ${\hat p}_\mu (\sigma_0) \Delta \sigma = {\hat p}_\mu =
- i \p/\p x^\mu (\sigma_0)$ is the partial derivative with respect to the
brane coordinates at $\sigma=\sigma_0 = 0$, and $\varphi (x^\mu) =
\phi (x^{\mu(\sigma)})|_{\sigma_0}$.

In our setup, the segment of the brane around $\sigma = \sigma_0 (=0)$
behaves as a point particle and satisfies the point-particle Klein-Gordon
equation. The remaining segment of the brane from $\sigma = \Delta \sigma$
to $\sigma = L$, has definite momentum $p_\mu (\sigma) = p_{0 \mu} (\sigma)$,
and contributes only a phase factor (\ref{3.2a}). If $p_{0 \mu} (\sigma) = 0$,
this means that actually there is no brane outside the range $\sigma \in (0,\Delta \sigma)$.
Then we have only the brane within $\sigma \in (0,\Delta \sigma)$, which in the
limit $\Delta \sigma \to 0$  behaves as a point particle. For finite, but small $\Delta \sigma$,
the brane behaves approximately as a point particle. At the end of Section 4 we further illuminate
the derivation of (\ref{3.8}) from (\ref{2.28}).

The action for the field $\varphi (x^\mu)$, satisfying the Klein-Gordon equation
(\ref{3.8}), is
\be
  I[\varphi (x^\mu)] = \frac{1}{2} \int \dd^D x \, (\p_\mu \vphi \p^\mu \vphi - m^2 \vphi^2).
\lbl{3.9}
\ee
The latter action can also be straightforwardly derived from the action (\ref{2.28})
by taking the ansatz
\be
  \phi(x^{\mu(\sigma)}) = {\rm e}^{\int_{\Delta \sigma}^L p_{0 \mu} (\sigma)
   x^\mu (\sigma) \dd \sigma} \vphi(x^\mu),
\lbl{3.10}
\ee
and using (\ref{3.3}).

In the following we will describe the flat brane by means of many particle non interacting field
theory. Different segments of the brane behave as distinguishable particles, each being
described by a different scalar field $\varphi_r (x)$. The action for a system of those scalar
fields is
\be
   I[\varphi_r (x)] = \frac{1}{2} \int \dd^D x \left ( \sum_{r=1}^N \p_\mu \vphi_r \p^\mu \vphi_r
   - m_r \vphi^2 \right ).
 \lbl{3.11a}
\ee
The canonically conjugated variables are $\vphi_r (t,{\bx})$ and $\Pi_r (t,{\bx}) =
\p {\cal L}/\p {\dot \vphi}_r = {\dot \vphi}_r$, where ${\bx} \equiv x^i$, $i=1,2,...,D-1$,
and the Hamiltonian is
\be
   H = \frac{1}{2} \int \dd^\bD {\bx} \sum_r (\Pi_r^2 - \p^i \vphi_r \p_i \vphi_r + m_r^2 \vphi_r^2)~,
   ~~~~i=1,2,3,...,\bD=D-1.
\lbl{3.11}
\ee

Upon quantization, $\vphi_r$ and $\Pi_r$ become the operators satisfying
\bear
   &&[\vphi_r(t,{\bx}),\Pi_s(t,{\bx}'] = i \delta^3 ({\bx}-{\bx'}) \delta_{rs}, \nonumber \\
   &&[\vphi_r(t,{\bx}),\vphi_s(t,{\bx'}] =0, ~~~~~[\Pi_r(t,{\bx}),\Pi_s(t,{\bx'}] =0.
\lbl{3.12}
\ear

The field $\vphi_r (x)$, $x \equiv x^\mu \equiv (t,{\bx})$ can be
expanded in terms of the creation and annihilations operators,
\be
 \vphi_r (x) = \int \frac{\dd^\bD{\bk}}
  {\sqrt{(2 \pi)^\bD 2 \omega_{\bk}}} (a_r ({\bk}) {\rm e}^{-ikx}
   + a_r^\dg ({\bk}) {\rm e}^{ikx}),
\lbl{3.12a}
\ee
satisfying
\be
  [a_r ({\bk}),a_s^\dg ({\bk'})] = \delta^\bD ({\bk}-{\bk}')
  \delta_{rs},
\lbl{3.13}
\ee
\be
   [a_r ({\bk}),a_s ({\bk'})] =0~,~~~~~
   [a_r^\dg ({\bk}),a_s^\dg ({\bk'})] =0.
\lbl{3.14}
\ee
We have absorbed the usual factor $(2 \pi)^\bD \,2 \omega_{\bk}$, where
$\omega_{\bk}=\sqrt{m^2+{\bk}^2}$, into the definition of operators
$a_r^\dg ({\bk})$, $a_r^\dg(\bk)$. The latter operators create and annihilate
a particle with the momentum ${\bk}$.

Let us introduce the Fourier transformed operators
\be
   a_r({\bx}) = \frac{1}{\sqrt{(2 \pi)^\bD}}\int \dd^\bD \bk \, a_r ({\bk})
                  {\rm e}^{i {\bk \bx}}~,~~~~~~~~~ 
  a_r^\dg({\bx}) = \frac{1}{\sqrt{(2 \pi)^\bD}}\int \dd^\bD \bk \, a_r^\dg ({\bk})
                    {\rm e}^{-i{\bk \bx}}
\lbl{3.15}
\ee
that satisfy
\be
  [a_r({\bx}),a_r^\dg({\bx'})]= \delta^\bD ({\bx}-{\bx'}) \delta_{rs}
\lbl{3.16}
\ee
\be
   [a_r ({\bx}),a_s ({\bx'})] =0~,~~~~~
   [a_r^\dg ({\bx}),a_s^\dg ({\bx'})] =0.
\lbl{3.17}
\ee
The operator $a_r ({\bx})$ annihilates the vacuum $|0\rangle$, whereas
$a_r^\dg({\bx})$ creates a particle at position ${\bx}$:
\be
  a_r^\dg({\bx})|0\rangle = |{\bx}\rangle.
\lbl{3.18}
\ee
In Appendix A we examine in more detail the properties of the operators
$a_r ({\bx})$, $a_r^\dg ({\bx})$ and show that in a given Lorentz frame
they can indeed be interpreted, respectively,  as a creation
and annihillaton operators for a particle at the position ${\bx}$.

A succession of $a_r^\dg ({\bx})$'s creates a manny particle state
\be
   a_1^\dg ({\bx}_1) a_2^\dg ({\bx}_2)...a_N^\dg ({\bx_N})|0\rangle
  = |{\bx}_1 {\bx}_2 ... {\bx}_N \rangle.
\lbl{3.19}
\ee
In a more compact notation this reads
\be
  \prod_r a_r^\dg ({\bx_r})\vac  \equiv A^\dg ({\bm X}_r)|0\rangle 
   = |{\bm X}_r \rangle~,~~~~r=1,2,...,N,
\lbl{3.20}
\ee
where ${\bm X}_r$ denotes a configuration of many particles, each having a different position ${\bx}_r$, 
$r=1,2,...,N$.

In the limit of infinitely many densely packed particles such a configuration
can be a brane:
\be
  \prod_\sigma a_\sigma^\dg ({\bx}_\sigma)|0\rangle \equiv
  A^\dg [{\bm X}(\sigma)]|0\rangle = |{\bm X}(\sigma)\rangle.
\lbl{3.30}
\ee

The momentum operator of the $r$-th particle is
\be
  {\hat {\bf p}}_r = \int \dd^\bD \bp\, a_r^\dg (\bp)\bp\, a_r (\bp)=
  \int \dd^\bD \bx \, a_r^\dg (\bx) (-i) \frac{\p}{\p \bx} \,a_r (\bx).
\lbl{3.31}
\ee
The latter definition is equivalent to the usual definition of momentum
operator, because the factor $1/((2 \pi)^\bD \,2 \omega_\bp)$ has been absorbed into
the definition of the operators $a_r^\dg (\bp)$ and $a_r (\bp)$.

Similarly, we can define the position operator,
\be
  {\hat {\bx}}_r = \int \dd^\bD \bx \, \bx \, a_r^\dg (\bx) a_r (\bx)
  = \int \dd^\bD \bp\, a_r^\dg (\bp) \,i \frac{\p}{\p \bp}a_r (\bp).
\lbl{3.31a}
\ee
Notice that the position operator so defined is not equivalent to the usually
defined ``position operator"\,\ci{PositionOperator}--\ci{PositionOperator3}, which is then shown to be inappropriate, because
it is not self-adjoint with respect to the considered, i.e., Lorentz invariant, scalar product.
Our position operator (\ref{3.31a}) is Hermitian, because ${\hat \bx}_r^\dg = {\hat \bx}_r$.
It is also self-adjoint with respect to the Lorentz non invariant\footnote{
We perform the integration over $\dd^\bD \bx$, where $\bD=D-1$,  i.e., over a $(D-1)$-dimensional hypersurface in
$D$-dimensional spacetime. For more details about the position operator so defined, see Appendices A--C.}
scalar product between the wave packet states created by $a_r^\dg (\bx)$ or, equivalently, by $a_r^\dg (\bp)$.

The commutator of those operators is
\be
   [\hbx_r,\hbp_s] = i \delta_{rs} {\bf 1} {\hat N}_r ~,~~~~~~{\bf 1} \equiv {\delta^i}_j~,~~~i,j=1,2,...D-1, ~~~D-1=\bD,
\lbl{3.32}
\ee
where ${\hat N}_r = \int \dd^\bD \bx \, a_r^\dg (\bx) a_r (\bx)$ is the number operator for an $r$-type
particle.

Let us define the {\it center of mass operator}
\be
   \hbx_{rT} \equiv {\hat N}_r^{-1} \hbx_r\
\lbl{3.33}
\ee
which satisfies
\be
   \hbx_{rT} \left (\prod_k a_r^\dg (\bx_k) \right )|0 \rangle = \bx_T  \left (\prod_k a_r^\dg (\bx_k) \right )|0 \rangle,
\lbl{3.34}
\ee
\be
  \bx_{rT} = \frac{1}{N} \sum_{k=1}^N \bx_{rk} .
\lbl{3.35}
\ee
Then Eq.\,(\ref{3.32}) can be written as
\be
  [\hbx_{rT},\hbp_s] = i \delta_{rs} {\bf 1}.
\lbl{3,36}
\ee

A generic many particle state is then a superposition
\bear
  &&|\psi \rangle = \int \dd \bx_1 \dd \bx_2 ...\dd \bx_N \, f(t,\bx_1,\bx_2,...,\bx_N)
  a_1^\dg (\bx_1) ...a_N^\dg (\bx_N)|0\rangle \nonumber\\
  &&\hs{6mm}=\int \dd \bp_1 \dd \bp_2 ...\dd \bp_N \, g(t,\bp_1,\bp_2,...,\bp_N)
  a_1^\dg (\bp_1) ...a_N^\dg (\bp_N)|0 \rangle .
\lbl{3.37}
\ear

Because $a_r^\dg (\bx)$, $r=1,2,...,N$, are bosonic operators, there can be more than one operator
of the same type $r$ in the product. Thus, $a_r^\dg (\bx_r)$ can be extended to 
$a_r^\dg (\bx_r)a_r^\dg (\bx'_r)...$, and $f(\bx_1,\bx_2,...,\bx_N)$ into
$f(\bx_1,\bx'_1,\bx''_1,...,$ $\bx_2,\bx'_2,\bx''_2,...,\bx_N,\bx'_N,\bx''_N,...)$. A superposition then
goes over all those possibilities. The wave function is symmetric with respect to the
interchange of  $\bx_1,\bx'_1,\bx''_1$, but has no definite symmetry with respect to the
interchange of $\bx_r$ and $\bx_{r'}$. The state (\ref{3.37}) contains coherent states as well.

A state evolves in time according to the Schr\"odinger equation
\be
  i \frac{|\Psi \rangle}{\p t} = H |\Psi \rangle,
\lbl{3.38}
\ee
where the Hamiltonian operator is given in Eq.\,(\ref{3.11}), which, after using (\ref{3.12})--(\ref{3.14})
becomes
\be
  H = \int \dd^\bD {\bk} \, \sum_r \omega_{r{\bk}} \left ( a_r^\dg ({\bk}) a_r ({\bk}) 
  + \frac{\delta (0)}{2} \right ).
\lbl{3.39}
\ee
Here $\omega_{r{\bk}} = \sqrt{m_r^2 + {\bk}^2}$. 

We are now interested in calculating the expectation value of the $r$-th particle position operator,
$\langle \hbx_{rT}\rangle$ in a state $|\psi \rangle$. After a straightforward calculation, by using
the Schr\"odinger equation, the commutation relations (\ref{3.13}),(\ref{3.14}),
(\ref{3.16}),(\ref{3.17}), the definition (\ref{3.31a}) of the position operator,
and by taking the Gaussian wave packet so that
$g(t,\bp_1,...,\bp_N) = g_1 (t,\bp_1) g_2 (t,\bp_2)...g_N (t,\bp_N)$,
with
\be
  g_r (t,\bp_r) = {\rm exp} \left [ -i (\omega_{r \bp} +E_0)t \right ]
   {\rm exp} {\frac{(\bp_r - \bp_{0r})^2 \sigma_0}{2}},
\lbl{3.39a}
\ee
where $E_0 = \frac{1}{2}\int \dd \bp \sum_r \omega_{r \bp}\, \delta(0)$,
we obtain
\be
  \langle \hbx_{rT} \rangle = \left \langle \frac{\hbp_r}{\omega_{r \bp}} \right \rangle t + \bx_{r 0} .
\lbl{3.40}
\ee
Here
\be
  \left \langle \frac{\hbp_r}{\omega_{r \bp}} \right \rangle \equiv
  \langle \psi | \frac{\hbp_r}{\omega_{r \bp}} |\psi \rangle =
  \int \dd \bp \frac{\bp \, t}{\omega_{r \bp}} g_r^* (\bp) g_r (\bp),
\lbl{3.41}
\ee
and $\bx_{r 0} = \langle \psi (t=0)| \hbx_r |\psi (t=0) \rangle$.

In the last equation we have an example for the expectation value
of the operator $\bp_r/\omega_{r \bp}$ in the case when for each $r$
we have only one particle state, $a_r^\dg (\bx)\vac$, and not
$a_r^\dg (\bx) a_r^\dg (\bx') a_r^\dg (\bx'') ... \vac$.

In Eq.\, (\ref{3.41}),  $\langle \hbp_r/\omega_{r \bp} \rangle$ is the expectation value
of a particle's velocity. Thus the expectation value of each particle's center of
mass position within our configuration traces a straight worldline. If particles
are close to each other, such a configuration samples a flat brane. In the continuous
limit we have a flat brane.

The position and momentum operator of the whole configuration are
\bear
  &&{\hat x}^{\bmu} = \sum_r  \int \dd^{\bar D} \bx \, x^\bmu \, a_r^\dg (\bx) a_r (\bx) \lbl{3.43}\\
  && {\hat p}^{\bmu} = \sum_r \int \dd^{\bar D} \bp \, p_\bmu \, a_r^\dg (\bp) a_r (\bp) \nonumber\\
  &&\hs{5mm} = \sum_r \int \dd^{\bar D} \bx \, \dd^\bD \bx' (-i) \p_\bmu \delta (\bx - \bx')  a_r^\dg (\bx) a_r (\bx')\nonumber\\
  &&\hs{5mm} = \sum_r \int \dd^\bD \bx\, a_r^\dg (\bx) (-i) \p_\bmu a_r(\bx) \lbl{3.44}
\ear
They satisfy
\be
  [{\hat x}^\bmu,{\hat p}_\bnu ]= i {\delta^\bmu}_\bnu \, {\hat N} ~, ~~~~~~\bmu, \bnu = 1,2,...\bD~, ~~~\bD=D-1,
\lbl{3.45}
\ee
where
\be
  {\hat N} = \sum_r \int \dd^\bD \bx \, a_r^\dg (\bx) a_r (\bx) = \sum_r {\hat N}_r
\lbl{3.46}
\ee
is the number operator for the whole configuration.

A single state of the $s$-the particle is
\be
  a_s^\dg (\bp) \vac = |\bp,s \rangle~,~~~~a_s^\dg (\bx) \vac = |\bx,s \rangle.
\lbl{3.47}
\ee
The matrix elements are
\bear
\langle \bp,s| \hbp |\hbp',s' \rangle = \bp \, \delta_{s s'} \, \delta (\bp - \bp')\nonumber\\
\langle \bx,s| \hbx |\hbx',s' \rangle = \bx\, \delta_{s s'}\, \delta (\bx - \bx')
\lbl{3.48}
\ear

All equations (\ref{3.11a})--(\ref{3.48}) can be straightoforwardly generalized to a
continuous set of ``particles", if instead of the discrete index $r$ we take
a continuous parameter, more precisely, a set of parameters $\sg\equiv \sg^\ba$,
$\ba = 1,2,...,p$.

\section{Towards curved brane space: A brane as a bunch of interacting point
particles}

Let us now introduce an interaction between the particles described by the fields
$\varphi_r$, and generalize the action (\ref{3.11a}) according to
\be
   I[\varphi^i] = \frac{1}{2} \int \dd^D x \, (\p_\mu \varphi^r \p^\mu \varphi^s - m^2 \vphi^r \vphi^s)s_{rs}.
\lbl{4.1}
\ee
The matrix $s_{rs}$ has the r\^ole of a metric in the space of fields. In general, $s_{rs}$ is
a functional of $\vphi^r$. If $s_{rs}$ is not a functional of $\vphi^r$, if it can be diagonalized,
and has the inverse $s^{rs}$, then the action (\ref{4.1}) brings nothing new in comparison
with the action (\ref{3.11a}). Interactions come into the game, if $s_{rs}$ is a functional of
$\vphi^r$, or if it cannot be diagonalized. 

In the continuum limit, the discrete index $r$ becomes the continuous index $(\sigma)$,
and $\vphi^r$ becomes $\vphi^{(\sigma)}$. A discrete set of point particles, described
by a discrete set of scalar fields $\vphi^r$, $r=1,2,...,N$, becomes a continuous set
of point particles ---a brane--- described by a continuous set of scalar fields $\vphi^{(\sigma)}$.
The action (\ref{4.1}) is then replaced by
\be
  I[\vphi^{(\sigma)}] = \frac{1}{2} \int \dd^D x \, \left ( \p_\mu \vphi^{(\sigma)} \p^\mu \vphi^{(\sigma'})
  - m^2 \vphi^{\sigma)} \vphi^{(\sigma')} \right ) s_{(\sigma)(\sigma')}.
\lbl{4.2}
\ee

In general, $s_{(\sigma)(\sigma')}$ is a functional of $\vphi^{(\sigma)}$, and it can thus
give an interaction, provided that the space of fields has nonvanishing curvature. We will
restrict our consideration to the case when $s_{(\sigma),(\sigma')}$ does not depend on
$\vphi^{(\sigma)}$. Then we can still have an interacting system, if $s_{(\sigma)(\sigma')}$
cannot be diagonalized to $s_{(\sigma)(\sigma')} = \delta (\sigma-\sigma')$. We will assume
that $s_{(\sigma)(\sigma')}$ has the inverse $s^{(\sigma)(\sigma')}$, such that
\be
  s^{(\sigma)(\sigma'')}s_{(\sigma'')(\sigma')} = {\delta^{(\sigma)}}_{(\sigma')} \equiv
  \delta (\sigma - \sigma').
\lbl{4.3}
\ee

The equation of motion is
\be
  \p_\mu \p^\mu \vphi_{(\sigma)} + m^2 \vphi_{(\sigma)} = 0,
\lbl{4.4}
\ee
where $\vphi_{(\sg)} = s_{(\sigma)(\sigma')} \vphi^{(\sigma')}$. Because of (\ref{4.3})
we also have
\be
  \p_\mu \p^\mu \vphi^{(\sigma)} + m^2 \vphi^{(\sigma)} = 0,
\lbl{4.5}
\ee

The canonically conjugated variables $\vphi^{(\sigma)}$ and $ \Pi_{(\sigma)} =
\p {\cal L}/\p {\dot \vphi}^{(\sigma)} = {\dot \vphi}_{(\sigma)}$ satisfy the commutation
relations
\be
  [\vphi^{(\sigma)} (x),\Pi_{(\sigma')} (x')]\Bigl\vert_{x^0=x'^0} = {\delta^{(\sigma)}}_{(\sigma')}
  \delta^\bD (\bx-\bx')
\lbl{4.6}
\ee
\be
  [\vphi^{(\sigma)} (x),\vphi^{(\sigma')} (x')] \Bigl\vert_{x^0=x'^0} =0~,~~~~~~~~~~~
   [\Pi_{(\sigma)} (x),\Pi_{(\sigma')} (x')]\Bigl\vert_{x^0=x'^0} =0 .
\lbl{4.7}
\ee

The Hamiltonian is
\be
  H= \int \dd^\bD \bx \, (\Pi_{(\sigma)} {\dot \vphi}^{(\sigma)} - {\cal L} ) =
  \frac{1}{2} \int \dd^\bD \bx \, (\Pi_{\sigma)} \Pi^{(\sigma)} - \p_i \vphi^{(\sigma)} \p^i \vphi_{(\sigma)}
  + m^2 \vphi^{(\sg)} \vphi_{(\sg)}).
\lbl{4.8}
\ee

A general solution of Eq.\,(\ref{4.4}) can be expanded according to
\be
  \vphi_{(\sg)} (x) = \int \frac{\dd^\bD \bk} {\sqrt{(2 \pi)^\bD 2 \omega_\bk}}
  \left ( a_{(\sg)} (\bk) {\rm e}^{-i k x} + a_{(\sg)}^\dg  (\bk) {\rm e}^{i k x} \right ) ,
\lbl{4.9}
\ee
where $\omega_\bk = \sqrt{\bk^2 + m^2}$.
Analogous expansion holds for $\vphi^{(\sg)}$.
Now we have the following commutation relations:
\be
  [a^{(\sg)}(\bp), a_{(\sg')}^\dg (\bp')] = {\delta^{(\sg)}}_{(\sg')} \delta^\bD (\bp - \bp').
\lbl{4.9a}
\ee
Because $a_{(\sg)} = s_{(\sg)(\sg')} a^{(\sg')}$, and $a^{(\sg)} = s^{(\sg)(\sg')} a_{(\sg')}$,
we also have
\be
  [a_{(\sg)}(\bp), a_{(\sg')}^\dg (\bp')] = s_{(\sg)(\sg')} \delta^\bD (\bp - \bp').
\lbl{4.9b}
\ee
and
 \be
  [a^{(\sg)}(\bp), a^{\dg (\sg')}(\bp')] = s^{(\sg)(\sg')} \delta^\bD (\bp - \bp').
\lbl{4.9c}
\ee 

Using (\ref{4.9}) and (\ref{4.9a}), the Hamiltonian (\ref{4.8}) becomes
\bear
  &&H = \frac{1}{2} \int \dd^\bD \bk \, \omega_\bk \left ( a_{(\sg)}^\dg (\bk) a^{(\sg)}(\bk)
  + a^{(\sg)} (\bk) a_{(\sg)}^\dg (\bk) \right ) \nonumber \\
  &&\hs{5mm}=  \int \dd^\bD \bk \, \omega_\bk \, a_{(\sg)}^\dg (\bk) a^{(\sg)} (\bk)+ H_{\rm z.p.},
\lbl{4.10}
\ear
where $H_{\rm z.p.}$ is the "zero point" Hamiltonian, and
\be
  a_{(\sg)}^\dg (\bk) a^{(\sg)}(\bk) = a^{\dg (\sg)} (\bk) a^{(\sg')} (\bk) s_{(\sg)(\sg')}
   =a_{(\sg)} ^\dg(\bk) a_{(\sg')} (\bk) s^{(\sg)(\sg')}.
\lbl{4.11}
\ee

For the momentum operator we obtain\footnote{
From the action (\ref{4.2}), using the standard field theoretic methods, we obtain the
stress-energy tensor 
${T^\mu}_\nu = \frac{\p {\cal L}}{\p \p_\mu \vphi^(\sg)} \p_\nu \vphi^{(\sg)} -
{\cal L} {\delta^\mu}_\nu$, and the momentum $P_\nu = \int \dd \Sigma_\mu {T^\mu}_\nu$.
Its spatial components are $P_\bmu = \int \dd^\bD \bx \, \vphi_{(\sg)} \p_\bmu \vphi^{(\sg)}$,
where we have taken the reference frame in which the hypersurface has
components $\dd \Sigma_\mu = (\dd \Sigma_0,0,0,...,0)$ with $\dd \Sigma_0 = \dd^\bD \bx$.
The Fourier transform of the integrand in $P_\bmu \equiv {\bm P}$ gives after quantization the
momentum operator (\ref{4.12}).
}
\bear
   &&{\hat \bp} = \frac{1}{2} \int \dd^\bD \bk \, \bk \left ( a_{(\sg)}^\dg (\bk) a^{(\sg)}(\bk)
  + a^{(\sg)} (\bk) a_{(\sg)}^\dg (\bk) \right ) \nonumber \\
 &&\hs{5mm} =  \int \dd^\bD \bk \,\bk \, a_{(\sg)}^\dg (\bk) a^{(\sg)} (\bk)+ {\hat \bp}_{\rm z.p.}.
 \lbl{4.12}
\ear

We will now assume that in general  $m$ depends on position $\sg$ on the brane\footnote{
In the discrete case this corresponds to each particle having a different mass $m_r$ (see
Eqs.\,(\ref{3.11a}), (\ref{3.11}), (\ref{3.39})).}.
Then also $\omega_{\bk} = \sqrt{m^2 + \bk^2}$ is function of $\sg$. We thus have
\be
  H= \int \dd \bk \, \omega_\bk (\sg) a^{\dg (\sg)} (\bk) a^{(\sg')} (\bk) s_{(\sg)(\sg')} + H_{\rm z.p.},
\lbl{4.13}
\ee
\be
  H^\dg= \int \dd \bk \, \omega_\bk (\sg') a^{\dg (\sg)} (\bk) a^{(\sg')} (\bk) s_{(\sg)(\sg')} + H_{\rm z.p.}.
\lbl{4.14}
\ee
In the expression for $H^\dg$ we have renamed $\sg \rightarrow \sg'$, $\sg' \rightarrow \sg$, and used
$s_{(\sg')(\sg)}=s_{(\sg)(\sg')}$. The Hamilton operator so modified is not Hermitian.  The momentum
operator remains unchanged and Hermitian.

Let us now calculate the time derivative of the expectation value of the momentum operator 
$\hbp$. We obtain:
\be
  \frac{\dd}{\dd t} \langle \psi|\hbp |\psi \rangle = \left ( \frac{\dd}{\dd t} \langle \psi | \right ) \hbp |\psi \rangle
  + \langle \psi |\hbp  \frac{\dd}{\dd t} |\psi \rangle = (-i)\langle \psi |\hbp H - H^\dg \hbp |\psi \rangle.
\lbl{4.15}
\ee
In the last step of the above equation we have used the Schr\"odinger equation and its hermitian conjugate,
\be
   i  \frac{\dd}{\dd t} |\psi \rangle = H |\psi \rangle~,~~~~~-i  \frac{\dd}{\dd t} \langle \psi | =\langle \psi  | H^\dg.
\lbl{4.16}
\ee
In quantum theory we have by definition
\be
  \frac{\dd}{\dd t} \langle \psi|\hbp |\psi \rangle = \langle \psi | \frac{\dd \hbp}{\dd t} |\psi \rangle.
\lbl{4.17}
\ee
Therefore, Eq.\,(\ref{4.15}) gives
\be
  \frac{\dd \hbp}{\dd t} = (-i) (\hbp H - H^\dg \hbp ).
\lbl{4.18}
\ee
The latter expression is equal to $(\dd \hbp/\dd t)^\dg$, therefore $(\dd \hbp/\dd t)$ is Hermitian,
as it should be. The zero point Hamiltonian, $H_{\rm z.p.} = H_{\rm z.p.}^\dg$, cancels out in Eq.\,(\ref{4.18}).

As the first step let us now consider the brane state which is the product of  "single particle" wave
packet profiles:
\be
~~~~~~~~~~~~~|\psi \rangle = \prod_\sg \int \dd^\bD \bp_{(\sg)} \, g^{(\sg)} (\bp_{(\sg)}) a_{(\sg)}^\dg (\bp_{(\sg)}) \vac ~~~~~~~~~ \mbox{\rm no integration over $(\sg)$}.
\lbl{4.19}
\ee
Acting on the latter state by an annihilation operator, we obtain
\be
  a^{(\sg')} (\bp'_{(\sg')}) |\psi \rangle = \int \dd \bp_{(\sg)} \dd \sg' \delta (\sg'-\sg)
   \delta (\bp'_{(\sg')}-\bp_{(\sg)}) g^{(\sg)} (\bp_{(\sg')})|{\bar \psi} \rangle 
   = g^{(\sg')} (\bp'_{(\sg')} |{\bar \psi} \rangle,
\lbl{4.20}
\ee
where $|\bar \psi \rangle$ is the product of all the single "particle" states. except the one picked up by
$a^{(\sg')} (\bp'_{(\sg')})$:
\be
  |{\bar \psi} \rangle = \left ( \prod_{\sg \neq \sg'} \int \dd \bp_{(\sg)} g^{(\sg)} (\bp_{(\sg)})
  a_{(\sg)}^\dg (\bp_{(\sg)}) \right ) \vac.
\lbl{4.21}
\ee
We thus have
\be
  \langle \psi | a^{\dg (\sg'')}(\bp''_{(\sg'')}) a^{(\sg')} (\bp'_{(\sg')}) |\psi \rangle =
  g^{*(\sg'')} (\bp''_{(\sg'')}) g^{(\sg')} (\bp'_{(\sg')}) \langle {\bar \psi}|{\bar \psi} \rangle,
\lbl{4.22}
\ee
where normalization can be such that $ \langle {\bar \psi}|{\bar \psi} \rangle =1$.

Alternatively, if we take the state
\be
  |\psi \rangle = \int \dd \bp \, g^{(\sg)} (\bp) a_{(\sg)}^\dg (\bp) \vac,
\lbl{4.23}
\ee
where
\be
  g^{(\sg)} (\bp) a_{(\sg)} (\bp) =g^{(\sg)} (\bp) a^{\dg(\sg')} (\bp) s_{(\sg)(\sg')},
\lbl{4.24}
\ee
and where now we have the integration over $(\sg)$, $(\sg')$, then we obtain
\be
  a^{(\sg')} (\bp') |\psi \rangle = g^{(\sg')} (\bp') \vac,
\lbl{4.25}
\ee
and
\be
  \langle \psi| a^{\dg (\sg'')} (\bp'') a^{(\sg')} (\bp') |\psi \rangle =
  g^{*(\sg'')} (\bp'') g^{(\sg')} (\bp') \langle 0| 0 \rangle,
\lbl{4.26}
\ee
where $\langle 0| 0 \rangle = 1$.

Comparison of Eqs.\,(\ref{4.20}),(\ref{4.22}) with (\ref{4.25}),(\ref{4.26}) reveals us that instead
of the state (\ref{4.19}) in which we have the product of the single "particle" states, we can as well
take the state (\ref{4.23}) in which we have a superposition of the single "particles" states over $\sg$.

Inserting the state (\ref{4.19}) into (\ref{4.15}), we obtain
\be
   \frac{\dd}{\dd t} \langle \psi|\hbp |\psi \rangle = (-i) \int \dd^\bD \bp \, \bp \, g^* (\sg,\bp) g(\sg',\bp)
   s(\sg,\sg') (\omega_\bp (\sg) - \omega_\bp (\sg')) \dd \sg \dd \sg' ,
\lbl{4.26a}
\ee
where we now write $g^{(\sg)} (\bp) \equiv g(\sigma,\bp)$, and $s_{(\sg)(\sg')} \equiv s(\sg,\sg')$.

Let us now choose
\be
  s_{(\sg)(\sg')} \equiv s(\sg,\sg') = (1 + \lambda \p_\ba \p^\ba)\, \delta^p (\sg - \sg').
\lbl{4.27}
\ee
Then Eq.\,(\ref{4.26a}) gives
\be
   \frac{\dd}{\dd t}\langle \bp \rangle \equiv  \frac{\dd}{\dd t} \langle \psi|\hbp |\psi \rangle 
   =(-i) \lambda  \int \dd^\bD \bp \, \dd^p \sg \, \bp \,\omega_\bp (\sg) (g^* \p_\ba \p^\ba g - \p_\ba \p^\ba g^* g ).
\lbl{4.28}
\ee
This is the time derivative of the expectation value of the total momentum operator of the brane,
\bear
  \langle \psi|\hbp |\psi \rangle &=& \int \dd \bp \, \dd \sg \, \dd \sigma' \, \bp \,g^*(\sg,\bp) g(\sg',\bp) s(\sg,\sg')
  \nonumber \\
  &=& \int \dd \bp \, \dd \sg \, \bp (g^* g + g^* \p_\ba \p^\ba g ) 
  = \langle \hbp \rangle = \int \dd \sg \langle \hbp \rangle_\sg
\lbl{4.29}
\ear
where
\be
  \langle \hbp \rangle_\sg = \int \dd \bp  \, \bp (g^* g + g^* \p_\ba \p^\ba g ) = \langle \psi|_\sg \hbp |\psi \rangle_\sg
\lbl{4.30}
\ee 
Here
\be
  |\psi \rangle_\sg = \int \dd \bp \, g(\sg,\bp) a^\dg (\sg,\bp) \vac ,
\lbl{4.31}
\ee
is the state of the brane's element at $\sg' \equiv \sg'^\ba$. This is the state (\ref{4.19}) in which there is
no product over $\sg$, or equivalently, the state (\ref{4.23}), in which there is no integration over $\sg$.
In Eq.\,(\ref{4.30}) we have thus the expectation value of the momentum of the brane's element at $\sg$.

Omitting the integration over $\sg$ in Eq.\,(\ref{4.28}), we obtain
\be
   \frac{\dd}{\dd t}\langle \hbp \rangle_\sg 
   =(-i) \lambda  \int \dd^\bD \bp \,\bp \,\omega_\bp (\sg) (g^* \p_\ba \p^\ba g - \p_\ba \p^\ba g^* g ),
\lbl{4.31a}
\ee
which is the time derivative of the expected momentum of a brane's element, i.e., a ``particle" forming the
brane.

In Eq.\,(\ref{4.31a}) we can take $\omega_\bp$ that does not change with $\sg$, and yet, in general,
the expression would not vanish. For constant $\omega_\bp$ we obtain
\be
   \frac{\dd}{\dd t}\langle \hbp \rangle_\sg 
   =(-i) \lambda \p_\ba \int \dd^\bD \bp \, \bp \, \omega_\bp (g^*  \p^\ba g -  \p^\ba g^* g ).
\lbl{4.31b}
\ee
This is the continuity equation for a current density on the brane. Integrating the latter equation over $\sg$,
we have $(\dd /\dd t) \langle \hbp \rangle = 0$. This means that for a constant $m$, and
thus for constant $\omega_\bp = \sqrt{m^2 + \bp^2}$, the total momentum of the brane is constant in
time, as it should be for an isolated brane\footnote{
The quantity $m$ is related to the brane tension $\kappa$ according to $m= \kappa \Delta \sg$
(see Eq.\,(\ref{3.7}) and the explanation above it).
If the tension $\kappa$, and thus $m$,  is not constant, this means that the brane is not isolated,
i.e., it is in an interaction with other physical systems, 
therefore the brane's total momentum changes with time.}.
The momentum of a brane's element $\dd \sg$ at $\sg$ in
general changes with time according to Eq.\,(\ref{4.31b}). With our model we have thus reproduced the
well known facts about the brane's momentum. In the following we will explore how the things look
in the coordinate representation.

By taking the Fourier transform according to
\be
  g(\sg,\bp) = \frac{1}{(2 \pi)^{\bD/2}} \int {\rm e}^{- i \bp \bx} f(\sg,\bx) \dd \bx,
\lbl{4.32}
\ee
we obtain
\be
  \frac{\dd}{\dd t}\langle \hbp \rangle_\sg 
  = - \lambda \p_\ba \int \dd^\bD \bx \, \left [ f^* (\sg,\bx) \left ( \nabla (-i)\frac{\p}{\p t}\p^\ba f(\sg,\bx) \right )
  -   \left ( \nabla (-i) \frac{\p}{\p t} \p^\ba f^*(\sg,\bx) \right )  f(\sg,\bx) \right ]
\lbl{4.33}
\ee
In the above equation, $- i \frac{\p}{\p t}$ comes from the Fourier transform of $\omega_\bp = \sqrt{m^2 + \bp^2}$, which gives
\be
(m^2 + (-i)^2 \nabla^2)^{1/2} f = - i \frac{\p}{\p t} f.
\lbl{4.34}
\ee
The latter equality comes from the Schr\"odinger equation, as shown in Appendix B.
Such equation is well known in the literature\,\ci{RelatSchEquation}--\ci{RelatSchEquation3},\ci{Al-Hashimi}.

Eq.\,(\ref{4.33}) can be written as
\be
  \frac{\dd}{\dd t}\langle {\hat p}_\bmu  \rangle_\sg 
    = - \lambda \p_\ba \int \dd^\bD \bx \, \left [ f^* (\sg,\bx) \left ( -i \frac{\p}{\p t} \, \p^\ba \p_\bmu f \right )
    -  \left ( -i \frac{\p}{\p t} \, \p^\ba \p_\bmu f^* \right ) f \right ],
\lbl{4.35}
\ee
where $\nabla \equiv \p_\bmu$, $\bmu = 1,2,...\bD$. The r.h.s. of Eq.\,(\ref{4.35}) is the divergence of the
expectation value of the operator
\be
   {{\hat \pi}^\ba}_{~\bmu} = -i \lambda \frac{\p}{\p t}{\stackrel{\leftrightarrow}{\p}}^{\ba} \p_\bmu
  \equiv -i \lambda  \frac{\p}{\p t} \left ( {\stackrel{\leftarrow}{\p}}^{\,\ba}  
  - {\stackrel{\rightarrow}{\p}}^{\,\ba}  \right ) \p_\bmu,
\lbl{4.35a}
\ee
which roughly corresponds to the classical quantity
\be
  {\pi^\ba}_\bmu = \kappa \sqrt{{\dot X}^2}\, \p^\ba X_\bmu = {\dot X}^2 p_0(\sg) \p^\ba X_\bmu , ~~~~{\rm where}
  ~~~p_0(\sg) = \frac{\kappa} {\sqrt{{\dot X}^2}}~,~~~X^0 \equiv t = \tau ,
\lbl{4.36}
\ee
associated with a brane for which the determinant  $\bar \gam$ of the spatial induced metric is constant, the constant
being equal to 1 (see Eq.\,(\ref{2.16ge})).
Such a brane can be either a {\it flat brane}, or a non flat brane, described in a gauge in which $-{\bar \gam}=1$.
The correspondence is approximate (rough), because in Eq.\,(\ref{4.35}) $\p^\ba$ is raised with $\delta^{\ba \bb}$,
and not with ${\bar \gam}^{\ba \bb}$.

In order to calculate the integral in Eq.\,(\ref{4.35}) we need to know the wave function  $f(\sg,\bx,t)$ which, in general
depends on time, and must satisfy the Schr\"odinger equation. Its exact solution for a minimal uncertainty
wave packet has been found in Ref.\,\ci{Al-Hashimi}. As a first estimation let us take the unperturbed wave function, which close to the initial time $t=0$, is equal to
\be
  f \approx A {\rm e}^{- \frac{(\bx-{\bar {\bm X}}(\sg))^2}{2 {\tl \sg}_0}} {\rm e}^{i {\bar \bp} \bx} {\rm e}^{i {\bar p}_0 t},
\lbl{4.37}
\ee
where ${\bar {\bx}}(\sg)$, ${\bar \bp}$, ${\bar p}_0$ are, respectively,  the coordinates, momentum and energy
of the wave packet center, and where $A$ is the normalization constant. Because $f$ satisfies Eq.\,(\ref{4.34}),
in which the square root expands to infinite order derivatives, at any $t > 0$, the function $f$, even if localized
at $t=0$, becomes delocalized. But as it follows from Ref.\,\ci{Al-Hashimi}, the deviation from the Gaussian
wave packet, such as (\ref{4.37}), is relatively small close to the initial time $t=0$. 

So we have
\be
  -i \frac{\p f}{\p t} \biggl\vert_{t=0} = {\bar p}_0 f~, ~~~~~~
  f(\sg,\bx,0) = A {\rm e}^{- \frac{(\bx-{\bar {\bm X}}(\sg))^2}{2 {\tl \sg}_0}} {\rm e}^{i {\bar \bp} \bx} 
\lbl{4.37a}
\ee
Introducing
\be
  \bx-{\bar {\bm X}}(\sg) = x^\bmu - {\bar X}^\bmu (\sg) = u^\bmu,
\lbl{4.39}
\ee
we have
\be
  \p_\ba f = \frac{u_\bnu \p_\ba {\bar X}^\bnu}{\sg_0} f~, ~~~~~~~\p_\bmu f = - \frac{u_\bmu}{\sg_0} f,
\lbl{4.40}
\ee
\be
  \p_\ba \p_\bmu f =  \left ( \frac{\p_\ba {\bar X}_\bmu}{\sg_0}  - \frac{u_\bmu u_\bnu \p_\ba {\bar X}^\bnu}{\sg_0^2}
  + i \p_\ba {\bar p}_\bnu u^\bnu + i \p_\ba p_0 t \right ) f,
\lbl{4.41}
\ee
\be
  \int \dd^\bD {\bx}\, f^* u_\bmu u_\bnu f = A^2 (\pi \sg_0)^{\bD/2} \frac{\sg_0}{2} \, \delta_{\bmu \bnu} ,  
\lbl{4.42}
\ee
\be
    \int \dd^\bD {\bx}\, f^* u_\bmu f = 0~,~~~~~ \int \dd^\bD {\bx}\, f^* u_\bmu u_\bnu u_{\bar \rho} f =0.
\lbl{4.43}
\ee
The normalization of $f$ involves also the integration over $\dd^p \sg$, so that we have
\be
  \int f^* f \dd^\bD x \dd^p \sg = A^2 (\pi \sg_0)^{\bD/2} S = 1 ~,~~~~~~S= \int \dd^p \sg .
\lbl{4.43a}
\ee
Using (\ref{4.37})--(\ref{4.43}) in Eq.\,(\ref{4.35a}) and taking $t \approx 0$, we obtain
\be
  \frac{\langle {\hat p}_\bmu \rangle_\sg}{\dd t}
   = - \lambda \p_\ba \left ( \frac{{\bar p}_0}{S}  \frac{\p^\ba {\bar X}_\bmu}{\sg_o} \right ) .
\lbl{4.44}
\ee

Let us compare the latter equations with the brane equation of motion (\ref{2.16ge}) in which we take 
$(-{\bar \gam})=1$. The expectation
value $\langle {\hat p}_\bmu \rangle_\sg \equiv \langle {\hat \bp} \rangle_\sg$ corresponds to
$p_\bmu (\sg) = \frac{\kappa {\dot X}_\bmu}{\sqrt{{\dot X}^2}}$, whereas $\frac{{\bar p}_0}{S} \p^\ba {\bar X}_\bmu$
corresponds to $\kappa \sqrt{ {\dot X}^2} \p^\ba X_\bmu 
= p_0 (\sg) {\dot X}^2 \p^\ba X_\bmu$. In the gauge $\tau = t \equiv X^0$,
the latter expression becomes
$p_0 (\sg)(1 - v^2) \p^\ba X_\bmu \approx p_0 (\sg) \p^\ba X_\bmu$, if $v^2 \approx 0$.
We have thus found that the centroid coordinates ${\bar X}_\bmu (\sg)$ satisfy the equations of
motion of a brane with $(-{\bar \gam})=1$ and $v^2 \approx 0$, up to the factor $\lambda/\sg_0$.
It is fascinating that such result comes from the field theory of a continuum of points particles, in which the interaction
is given in terms of the metric (\ref{4.27}) acting in the space of fields $\phi_{(\sg)} (t,\bx)$, and the wave
packet profile being approximated with the expression (\ref{4.37})  taken near $t=0$.
Therefore Eq.\,(\ref{4.44}) is valid only near the initial time. For a different quantum state we
would obtain  an equation of motion for the expectation values that would differ from (\ref{4.44}). 

We can also consider the possibility of introducing a more general interaction than (\ref{4.27}).
First we observe that Eq.\,(\ref{4.33}) can be written in the form
\be
  \frac{\dd}{\dd t} \langle \hbp \rangle_\sg = - \int \dd \sg' \dd^\bD \bx (-i) \left [ f^*(\sg,\bx)( \p_t \nabla f(\sg',\bx))
   - (\p_t \nabla f^* (\sg,\bx)) f \right ] \lambda (\sg,\sg')
\lbl{4.45}
\ee
where
\be
  \lambda(\sg,\sg') \equiv \lambda_{(\sg)(\sg')} = \lambda\, \p_\ba \p^\ba \delta (\sg - \sg').
\lbl{4.46}
\ee
If we generalize the interaction metric $\lambda(\sg,\sg')$ according to
\be
  \lambda(\sg,\sg') = \lambda \,\p'_\ba \left ( \sqrt{-{\bar \gam}(\sg')} \p^\ba \right ) \delta (\sg - \sg') ,
\lbl{4.47}
\ee
where ${\bar \gam} = {\rm det} \gam_{\ba \bb}$ is the determinant of the metric
 $\gam_{\ba \bb}$  in the space of parameters $\sg \equiv \sg^\ba$, and $\p^\ba = \gam^{\ba \bb} \p_\bb$,
 then we obtain
\be
   \frac{\dd}{\dd t} \langle \hbp \rangle_\sg 
   = -\lambda \p_\ba \int \dd^\bD \bx  \sqrt{-{\bar \gam}}\,  (-i) \left [ f^* (\p_t \p^\ba {\nabla} f) - (\p_t \p^\ba \nabla f^*) f \right ]
\lbl{4.48}
\ee
The latter equation corresponds to the brane equation of motion with non trivial $-{\bar \gam}\neq 1$,
i.e., to the equation of motion of the Dirac-Nambu-Goto brane, provided that $\gam_{\ba \bb}$ is equated with
the induced metric on the brane's worldsheet, $\gam_{\ba \bb}= \p_\ba X^\mu \p_\bb X^\mu$.

Taking the appropriate wave packet (\ref{4.37}) and performing the calculations as in (\ref{4.37a})--(\ref{4.43a})
we obtain
\be
  \frac{\langle {\hat p}_\bmu \rangle_\sg}{\dd t}
   = - \lambda \p_\ba \left ( \frac{{\bar p}_0}{S}  \frac{\sqrt{-{\bar \gam}} \gam^{\ba \bb}\p_\bb {\bar X}_\bmu}{\sg_o} \right ) .
\lbl{4.48a}
\ee
This is indeed very close to the brane equation (\ref{2.16g}) or (\ref{2.16ge}), apart from the factor
${\dot X}^2 = 1- v^2$. In our equation (\ref{4.48a}) we have ${\dot X}^2 =1$, which means that
$v^2 = 0$. Since Eq.\,(\ref{4.48a}) has been calculated for the wave packet at $t \approx 0$,
this is consistent with vanishing $\langle {\hat p}_\bmu \rangle \propto {\dot {\bar X}_\bmu}$ at $t \approx 0$.
Because by our assumption ${\bar \gam}_{\ba \bb}$ is the metric in the space of parameters
$\sg \equiv \sg^\ba$, $a=1,2,...,p$, and because ${\bar X}^\bmu (\sg)$ describe a brane whose induced
metric is $\p_\ba X^\bmu \p_\bb X_\bmu$, we conclude that
${\bar \gam}_{\ba \bb} =\p_\ba X^\bmu \p_\bb X_\bmu$.

Let us now investigate what happens if we use the metric (\ref{4.27}) in the classical action (\ref{2.8b}) by setting
\be
  \rho_{\mu(\sg) \nu(\sg')} = \eta_{\mu \nu} s_{(\sg)(\sg')} = \eta_{\mu \nu} (1+\p_\ba \p^\ba )\,  \delta (\sg - \sg')
\lbl{4.49}
\ee
Because the latter metric does not functionally depend on $X^{\alpha(\sg)}$, the second term in the
equation of motion (\ref{2.16a}) vanishes. Therefore the equation of motion is
\be
  \frac{\dd}{\dd \tau} p_{\mu(\sg)} = 0,
\lbl{4.50}
\ee
where
\be
  p_{\mu(\sg)} = \rho_{\mu(\sg)\nu(\sg')} p^{\nu(\sg')} =
 \frac{{\tl \kappa} {\dot X}_{\mu(\sg)}}{({\dot X}^{\nu(\sg')} {\dot X}_{\nu (\sg')})^{1/2}}
\lbl{4.51}
\ee
For the metric (\ref{4.49}) we have
\bear
  p_{\mu(\sg)} &=& \frac{{\tl \kappa} {\dot X}_\mu} { \sqrt{{\dot {\tl X}}^2}}
  + \lambda \p_\ba \left (  \frac{{\tl \kappa} \,\p^\ba X_\mu } { \sqrt{{\dot {\tl X}}^2}}    \right ), \nonumber \\
  &=& \frac{{\kappa} {\dot X}_\mu} { \sqrt{{\dot {X}}^2}}
  + \lambda \p_\ba \left (  \frac{{\kappa} \,\p^\ba X_\mu } { \sqrt{{\dot {X}}^2}}    \right ),
\lbl{4.51a}
\ear
where in the last step we have used Eq.\,(\ref{I6}) with $(-{\bar \gam})=1$.

We see that our metric (\ref{4.49}) modifies the momentum so that it contains an extra term, but otherwise
the equation of motion is merely the derivative of momentum (\ref{4.50}), with no brane-like
``force" term of the form similar to the second term in Eq.\,(\ref{2.16ge}).

For a generic metric, the constraints (\ref{I3}), associated with the action (\ref{2.8b}) leads to the field theory based
on the action
\be
  I[\phi] = \frac{1}{2} \int {\cal D} x \left ( \rho^{\mu(\sg) \nu(\sg')} \p_{\mu(\sg)} \phi \p_{\nu(\sg')} \phi
  - {\tl \kappa}^2 \phi^2 \right ).
\lbl{4.52}
\ee
I we take the Ansatz
\be
  \phi(x^{\mu(\sg)}) = \prod_{\sg''} \varphi_{(\sg'')} (x_{(\sg'')}^\mu),
\lbl{4.53}
\ee
then the functional derivative acts as a partial derivative according to
\be
  \p_{\mu(\sg)} \phi = \lim_{\Delta \sg \to 0} \frac{1}{\Delta \sg} \frac{\p_\mu \varphi_{(\sg)}}{\p x_{(\sg)}^\mu}
  \prod_{\sg'' \neq \sg} \varphi_{(\sg'')} (x_{(\sg'')}^\mu).
\lbl{4.54}
\ee
The second term in (\ref{4.52}) can be written in the form\footnote{This comes from a massless action in higher
dimensions.}
\be
  {\tl \kappa}^2 \phi^2= s^{(\sg)(\sg')} \kappa_{(\sg)} \kappa_{(\sg')} \phi^2 .
\lbl{4.55}
\ee

If we take the metric
\be
  \rho^{\mu(\sg)\nu(\sg')} = \eta^{\mu \nu} s^{(\sg)(\sg')},
\lbl{4.56}
\ee
where $s^{(\sg)(\sg')}$ is the inverse of $s_{(\sg)(\sg')} = (1+ \lambda \p_\ba \p^\ba) \delta (\sg-\sg')$,
we arrive at the action
\bear
  I[\varphi] &=& \frac{1}{2} \int \dd^D x \, s^{(\sg)(\sg')} \left ( \eta^{\mu \nu} \p_\mu \varphi_{(\sg)} \p_\nu \varphi_{(\sg')}
  - m^2 \vphi_{(\sg)} \vphi_{(\sg')} \right ) \nonumber \\
  &=&  \frac{1}{2} \int \dd^D x \, s_{(\sg)(\sg')} \left ( \eta^{\mu \nu} \p_\mu \varphi^{(\sg)} \p_\nu \varphi^{(\sg')}
  - m^2 \vphi_{(\sg)} \vphi_{(\sg')} \right ) .
\lbl{4.57}
\ear
This is just the action (\ref{4.2}) considered at the beginning of this section, by postulating the interaction
metric $s_{(\sg)(\sg)}$ between the continuous set of scalar fields $\vphi^{(\sg)} (x)$, whose quantized
theory leads to the expectation value equations of motion (\ref{4.44}), which contain the brane-like
force term that is missing in the classical equations of motion (\ref{4.50}) for the metric
$\rho_{\mu(\sg) \nu(\sg')} = \eta_{\mu \nu} s_{(\sg) (\sg')}$. The important point is that in the classical
theory with the relatively simple metric (\ref{4.49}) we have the simple equations of motion (\ref{4.50}),
whereas in the quantized theory with the same metric we also obtain the ``force" term in the effective
equations of motion (\ref{4.44}) or (\ref{4.48a}). The expectation value equations of motion describe
the centroid brane whose brane space metric is no longer (\ref{4.49}), but a more general effective
metric.

We will now show that to Eq.\,(\ref{4.44}) corresponds the metric
\be
  \rho_{\mu(\sg) \nu(\sg')} = \p_\ba X^\bmu \p^\ba X_\bmu \delta(\sg-\sg').
\lbl{4.58}
\ee
If we insert this metric into (\ref{2.8b}), we obtain the following equations of motion:
\be
  \frac{\dd}{\dd \tau} \left ( \frac{{\tl \kappa} \p_\ba X^\bnu \p^\ba X_\bnu {\dot X}_\mu}{\sqrt{{\dot {\tl X}}^2}}  \right ) 
  + \p_\ba \left (  \frac{{\tl \kappa} {\dot X}^2 \p^\ba X_\mu}{\sqrt{{\dot {\tl X}}^2}} \right ) = 0 ,
\lbl{4.58a}
\ee
where $\sqrt{{\dot {\tl X}}^2}\equiv {\dot X}^{\mu(\sg)} {\dot X}_{\mu(\sg)}
 = \rho_{\mu(\sg) \nu(\sg')} {\dot X}^{\mu(\sg)} {\dot X}^{\nu(\sg')}$, and ${\dot X}^2 \equiv
 {\dot X}^\mu {\dot X}_\mu = \eta_{\mu \nu} {\dot X}^\mu {\dot X}^\nu$.

Let us now observe that the following relation is satisfied:
\be
  \frac{{\tl \kappa}}{\sqrt{{\dot {\tl X}}^2}} = \frac{\kappa}{\sqrt{{\dot X}^2} \sqrt{\p_\ba X^\bmu \p^\ba X_\bmu}} .
\lbl{4.59}
\ee
This relation can be easily proved by writing it in the form
\be
  {\tl \kappa}^2 \p_\ba X^\bmu \p^\ba X_\bmu {\dot X}^2 = \kappa^2 {\dot X}^{\mu(\sg)} {\dot X}_{\mu(\sg)} ,
\lbl{4.60}
\ee
and integrating by $\dd \sg$. Then we obtain
\be
  {\tl \kappa}^2 \int \dd \sg \,\p_\ba X^\bmu \p^\ba X_\bmu {\dot X}^2 
  = \left ( \int \kappa^2 \, \dd \sg \right ) {\dot X}^{\mu(\sg)} {\dot X}_{\mu(\sg)} ,
\lbl{4.61}
\ee
Since   $\int \kappa^2 \dd \sg = {\tl \kappa}^2$ and
$\int \dd \sg \,\p_\ba X^\bmu \p^\ba X_\bmu \, {\dot X}^\nu {\dot X}_\nu
= \rho_{\mu(\sg) \nu(\sg'')} {\dot X}^{\mu(\sg)} {\dot X}^{\nu(\sg')}$, where $\rho_{\mu(\sg) \nu(\sg')}$ is
given by (\ref{4.58}), we see that (\ref{4.61}) is an identity.

Using (\ref{4.59}) in Eq.\,(\ref{4.58a}), we have
\be
  \frac{\dd}{\dd \tau} \left ( \frac{{ \kappa} \sqrt{\p_\ba X^\bnu \p^\ba X_\bnu} {\dot X}_\mu}{\sqrt{{\dot {X}}^2}}  \right ) 
  + \p_\ba \left (  \frac{{\kappa} {\dot X}^2 \p^\ba X_\mu}{\sqrt{{\dot {X}}^2} \sqrt{\p_\ba X^\bnu \p^\ba X_\bnu}} \right ) = 0 .
\lbl{4.62}
\ee
If we rewrite the latter equation in terms of momenta
\be
  p_{\mu(\sg)} = \frac{{\tl \kappa} \p_\ba X^\bnu \p^\ba X_\bnu {\dot X}_\mu (\sg)}
  {({\dot X}^{\alpha(\sg)} {\dot X}_{\alpha(\sg)})^{1/2}} =
  \frac{\kappa \sqrt{\p_\ba X^\bnu \p^\ba X_\bnu} {\dot X}_\mu}{\sqrt{{\dot X}^2}}
\lbl{4.63}
\ee
\be
  p^{\mu(\sg)} = \frac{{\tl \kappa}  {\dot X}_\mu}
  {({\dot X}^{\alpha(\sg)} {\dot X}_{\alpha(\sg)})^{1/2}} =
  \frac{\kappa  {\dot X}^\mu}{\sqrt{\p_\ba X^\bnu \p^\ba X_\bnu}\sqrt{{\dot X}^2}} ,
\lbl{4.64}
\ee
we obtain
\be
  \frac{\dd p_{\mu(\sg)}} {\dd \tau} + \p_\ba (p^{0 (\sg)} {\dot X}^2 \,\p^\ba X_\mu ) = 0.
\lbl{4.65}
\ee
For the spatial components $\bmu  =1,2,...{\bar D}$, in the gauge $\tau = X^0 \equiv t$, so that
${\dot X}^2 \equiv {\dot X}^\mu {\dot X}_\mu = 1 - v^2$, the latter equation matches the
expectation value equation (\ref{4.44}) if $v^2=0$.

Similarly, the expectation value equation of motion (\ref{4.48a}) can be derived from the effective action
(\ref{2.8b}) with the metric
\be
  \rho_{\mu(\sg) \nu(\sg')} = \sqrt{-{\bar \gam}} (\gam^{\ba \bb} \p_\ba X^\bmu \p_\bb X_\bmu)^p
 \, \delta (\sg - \sg')  .
\lbl{4.66}
\ee
The equations of motion are then
\be
  \frac{\dd}{\dd \tau} \left ( \frac{{\tl \kappa} \sqrt{-{\bar \gam}} (\p_{\bar c} X^\bnu \p^{\bar c} X_\bnu)^p
   {\dot X}_\mu}{\sqrt{{\dot {\tl X}}^2}}  \right ) 
  + \p_\ba \left (  \frac{{\tl \kappa} \sqrt{-{\bar \gam}} {\dot X}^2 \, p (\p_{\bar c} X^\bnu \p^{\bar c} X_\bnu)^{p-1}
  \gam^{\ba \bb}\p_\bb X_\mu}{\sqrt{{\dot {\tl X}}^2}} \right ) = 0 .
\lbl{4.67}
\ee
Instead of (\ref{4.59}) we have now
\be
  \frac{{\tl \kappa}}{\sqrt{{\dot {\tl X}}^2}} = \frac{\kappa}{\sqrt{{\dot X}^2}\,
  (\p_\ba X^\bmu \p^\ba X_\bmu)^{p/2}} .
\lbl{4.68}
\ee
Using the latter relation, Eq.\,(\ref{4.67}) becomes
\be
  \frac{\dd}{\dd \tau} \left ( \frac{{\kappa} \sqrt{-{\bar \gam}} (\p_{\bar c} X^\bnu \p^{\bar c} X_\bnu)^{p/2}
   {\dot X}_\mu}{\sqrt{{\dot {X}}^2}}  \right ) 
  + \p_\ba \left (  \frac{{\kappa} \sqrt{-{\bar \gam}} {\dot X}^2 \, p (\p_{\bar c} X^\bnu \p^{\bar c} X_\bnu)^{\frac{p}{2}-1}
  \p^\ba X_\mu}{\sqrt{{\dot {X}}^2}} \right ) = 0 .
\lbl{4.69}
\ee
Recall that ${\bar \gam}_{\ba \bb}$ is the metric in the space of parameters $\sg^\ba$. On the other
hand, $X^\bnu (\tau,\sg^\ba)$ describes a brane. Therefore it makes sense to
equate ${\bar \gam}_{\ba \bb}$ with the induced metric on the brane:
\be
  {\bar \gam}_{\ba \bb} = \p_\ba X^\bnu \p_\bb X_\bnu .
\lbl{4.70}
\ee
Let us now take into account that the trace of the metric is equal to the dimension of the brane:
\be
   \p_\bb X^\bnu \p^\bb X_\bnu = p .
\lbl{4.71}
\ee
Inserting this into (\ref{4.69}) we find that $p$ cancels out, and Eq.\,(\ref{4.69}) becomes
\be
   \frac{\dd}{\dd \tau} \left ( \frac{{\kappa} \sqrt{-{\bar \gam}}    {\dot X}_\mu}{\sqrt{{\dot {X}}^2}}  \right ) 
  + \p_\ba \left (  \frac{{\kappa} \sqrt{-{\bar \gam}} {\dot X}^2 \,  \p^\ba X_\mu}{\sqrt{{\dot {X}}^2}} \right ) = 0 .
\lbl{4.72}
\ee  
This is precisely the equation of motion (\ref{2.16ge}) of the Dirac-Nambu-Goto brane.
Expressing it in terms of momenta, it can be written in the form
\be
  \frac{\dd p_{\mu(\sg)}}{\dd \tau} + \p_\ba \left ( p^{0(\sg)} \,{\dot X}^2 \, \gam^{\ba \bb} \p_\bb X_\mu \right ) = 0.
\lbl{4.73}
\ee
In the gauge $\tau=X^0 \equiv t$, we find that for $v^2=0$ the latter equation corresponds to the expectation value
equation of motion (\ref{4.48a}).

\section{Further clarification of the action for a continuous set of interacting fields}

We will now rewrite the action (\ref{4.2}) into a more familiar form. Let us write
$\vphi^{(\sg)} (x) \equiv \vphi(\sg,x)$, and take the metric (\ref{4.27}). Then, after performing
partial integration over $\sg^\ba$ and omitting the surface term, we obtain
\be
  I = \frac{1}{2}\int \dd^D x\, \dd^p \sg \left [ \p_\mu \vphi \p^\mu \vphi - m^2 \vphi^2 +
  \lambda (\p_\mu \p_\ba \vphi \p^\mu \p^\ba \vphi - m^2 \p_\ba \vphi \p^\ba \vphi)    \right ] .
\lbl{5.1}
\ee
Variation with respect to $\vphi(\sg,x)$ gives the field equation
\be
  (\p_\mu \p^\mu + m^2)(1 + \lambda \p_\ba \p^\ba ) \vphi - 0.
\lbl{5.2}
\ee

A particular solution is
\be
  \vphi = {\rm e}^{- i \pi_\ba \sg^\ba} . {\rm e}^{-i p_\mu x^\mu} ,
\lbl{5.3}
\ee
subjected to the condition
\be
  (p_\mu p^\mu - m^2) (1 - \lambda \pi_\ba \pi^\ba ) = 0.
\lbl{5.4}
\ee
We can take $\pi_\ba$ arbitrary, whereas $p_\mu$ satisfying the mass shell constraint
$p_\mu p^\mu - m^2 = 0$ (Case A), or vice versa, $p_\mu$ arbitrary and $\pi_\ba$ satisfying
$1- \lambda \pi^\ba \pi_\ba = 0$ (Case B).

Eq.\,(\ref{5.2}) can be written as
\be
  \int \dd^p \sg' (\p_\mu \p^\mu + m^2) s(\sg,\sg') \vphi (\sg',x) = 0 ,
\lbl{5.5}
\ee
where
\be
  s(\sg,\sg') = (1+ \lambda \p_\ba \p^\ba) \delta^p (\sg - \sg').
\lbl{5.6}
\ee
The Fourier transform\footnote{
For simplicity reasons we use here the same symbol for the Fourier transformed quantity, but with
different arguments.}
 of $s(\sg,\sg')$ is
\be
  s(\pi,\pi') = \delta^p (1 - \lambda \pi_\ba \pi^\ba),
\lbl{5.7}
\ee
whose inverse is
\be
  {\tl s} (\pi,\pi') = \frac{\delta^p (\pi - \pi')}{1 - \lambda \pi^\ba \pi_\ba} .
\lbl{5.7a}
\ee
Taking the Fourier transform of the latter expression, we obtain
\be
  {\tl s}(\sg,\sg') = \int \frac{\dd^p \pi\, {\rm e}^{i \pi_\ba (\sg^\ba - \sg'^\ba)}}{1 - \lambda \pi^\ba \pi_\ba} .
\lbl{5.8}
\ee
This is the propagator in the space of parameters $\sg^\ba$, and is
the inverse of $s(\sg,\sg')$. We thus have
\be
  \int \dd \sg'' {\tl s}(\sg,\sg'') s (\sg'',\sg') = \delta^p (\sg - \sg') .
\lbl{5.9}
\ee
Multiplying Eq.\,(\ref{5.5}) by ${\tl s}(\sg'',\sg)$, integrating over $\sg$, using (\ref{5.9}), and
renaming $\sg'$ into $\sg$, we obtain
\be
  (\p_\mu \p^\mu + m^2) \vphi(\sg, x) = 0 .
\lbl{5.10}
\ee
Eqs.\,(\ref{5.2}) and (\ref{5.10}), of course, correspond to   (\ref{4.4}) and (\ref{4.5}), where
\be
  (1 + \lambda \p_\ba \p^\ba) \vphi(\sg,x) = \chi(\sg,x) \equiv \vphi_{(\sg)} (x).
\lbl{5.11}
\ee
A general solution of Eq.\,(\ref{5.2}) in Case A is
\be
  \vphi(\sg,x) = \int \frac{\dd^\bD \bp \, \dd^p {\bm \pi}}{\sqrt{(2 \pi)^\bD 2 \omega_\bp}} \left [ a({\bm \pi},\bp) {\rm e}^{- i \pi_\ba \sg^\ba}
  {\rm e}^{- i p_\mu x^\mu} +  a^\dg({\bm \pi},\bp) {\rm e}^{i \pi_\ba \sg^\ba}
   {\rm e}^{i p_\mu x^\mu} \right ] .
\lbl{5.12}
\ee
This can be written as
\be
  \vphi(\sg,x) = \int \frac{\dd^\bD \bp}{\sqrt{(2 \pi)^\bD 2 \omega_\bp}} \left [ a(\sg,\bp) 
  {\rm e}^{- i p_\mu x^\mu} +  a^\dg(\sg,\bp) {\rm e}^{i p_\mu x^\mu} \right ] .
\lbl{5.13}
\ee
where
\be
  a(\sg,\bp) = \int \dd^p {\bm \pi} a({\bm \pi},\bp) {\rm e}^{- i \pi_\ba \sg^\ba} .
\lbl{5.14}
\ee
Identifying $a(\sg,\bp)\equiv a_{(\sg)} (\bp)$ we find that (\ref{5.13}) is the same
equation as (\ref{4.9}), as it should be.

We see that our continuous bunch of scalar fields whose mutual interaction is given
by the field space metric $s_{(\sg)(\sg')} \equiv s(\sg,\sg')$ (given in Eq.\,(\ref{4.27}) or (\ref{5.6}))
 is described by the action (\ref{5.1}). This is an action
for a field $\vphi(\sg^\ba, x^\mu)$, which depend not only on spacetime coordinates $x^\mu$,
but also on the brane parameters $\sg^\ba$.  The action can be written in terms of the metric (\ref{5.6}),
whose inverse is the {\it propagator} (\ref{5.8}) on the brane.

\section{Conclusion}

We have found a resolution of the problem of brane quantization, which can have important
implication for the brane world scenarios that consider our spacetime as a brane living in
a higher dimensional space.
First we have shown that the Dirac-Nambu-Goto brane can be described as a ``point particle"
in an infinite dimensional brane space with a special metric. The analogy with general relativity
suggests that the metric is dynamical and thus not necessarily restricted to the special form.
As in general relativity the simplest metric is that of flat spacetime, so in the brane theory
the brane space can have a simple ``flat" metric as well. A flat brane is like a bunch of
non interacting point particles. Upon quantization such a system is described by the
quantum field theory of a continuous set of non interacting fields $\vphi_\sg$, each one
describing a different distinguishable particle.

We then considered an interacting system by introducing a coupling between the fields.
We achieved this by adding an extra term to the $\delta$-function like metric
in the field space. This extra, interacting, term was of the form
$\lambda \p_\ba \p^\ba \delta^p (\sg-\sg')$. Because of the latter term, the time derivative of
the expectation value of the momentum operator, calculated for an evolving wave packet like
state, does not vanish. We have found that the center of the wave packet at each $\sg^\ba$
satisfies the equations of motion of a classical brane which is nearly like the usual Dirac-Nambu-Goto
brane. The difference is in the determinant ${\bar \gam} \equiv {\rm det} \gam_{\ba \bb}$ of the
induced metric on the brane being restricted to ${\bar \gam} = -1$. We also showed that the interacting
term $\lambda \sqrt{-{\bar \gam}} \p_\ba \gam^{\ba \bb} \p_\bb \delta^p (\sg-\sg')$ for a general metric
$\gam_{\ba \bb}$ and the corresponding determinant ${\bar \gam}$ leads to the equation of motion of the Dirac-Nambu-Goto brane.

All this means that the special brane space metric for a Dirac-Nambu-Goto brane was induced
from the underlying field theory of the continuous system of interacting scalar fields, the interaction
being given by a certain coupling term. If we chose a different coupling term, we would obtain a
different effective classical brane, living in a brane space whose metric were different from that
of the Dirac-Nambu-Goto brane. 

In this paper we concentrated on scalar field. But analogous procedure could be applied to
fermion fields as well. We considered the usual canonical field quantization,  which is somewhat
cumbersome because of the $(1+3)$ split of spacetime. How the present theory can be cast into the
more elegant Fock-Schwinger proper time formalism, or into the Stueckelberg invariant
evolution parameter formalism, is beyond the scope of this paper, and will
be shown elsewhere.

\vs{1cm}

{\bf \large Appendix A:  Position operator}

\vs{2mm}

The creation operator
$a^\dg (\bx)$ for a particle at position $\bx$, although not Lorentz covariant, is not problematic,
if it is understood that its definition (\ref{3.15}) is valid only in a given Lorentz reference frame.
Namely, in the quantum field theory of a scalar field we have the operator $a^\dg (\bp)$ that
creates a particle with momentum $\bp$. The Fourier transform of $a^\dg (\bp)$ is
\be
   a^\dg (\bx) = \frac{1}{\sqrt{(2 \pi)^3}} \int \dd^3 \bk \,a^\dg (\bk) {\rm e}^{- i \bk \bx}
\lbl{3.15a}
\ee
The position operator is then
\be
  {\hat \bx} = \int \dd^3 \bx \,a^\dg (\bx)\, \bx \,a(\bx) = \int \dd^3 \bp \, a^\dg (\bp)\, i \frac{\p}{\p \bp} a(\bp) .
\lbl{A1}
\ee
This is quite a legitimate operator in the Lorentz frame with respect to which
the time $t \equiv x^0$ and space $\bx \equiv x^\bmu$, $\bmu=1,2,3$ are defined.
The difference with the usual treatment is that our position
creation and annihilation operators $a^\dg (\bx)$, $a(\bx)$ are not identified with the
field operators $\vphi (x)$ (or the positive or negative frequency part of $\vphi (\bx)$). They are
just Fourier transforms of $a^\dg(\bp)$ and $a(\bp)$. Since the latter operators are legitimate objects
of the field theory, also $a^\dg (\bx)$, $a(\bx)$ are legitimate objects in a given Lorentz
frame $S$, although they are not covariant objects. In a different
Lorentz frame one must define those operators anew. 

Taking the Hamiltonian\footnote{We omit the zero point term.}
\be
  H = \int \dd^3 \bp \, \omega (\bp) a^\dg (\bp) a(\bp) = {\hat p}^0 ,
\lbl{A2}
\ee
we have
\be
  {\dot {\hat \bx}} = - i [{\hat \bx},H] = \int \dd^3 \bp \, a^\dg (\bp) a (\bp) \frac{\p \omega}{\p \bp}
  = \int \dd^3 \bp \, a^\dg (\bp) a(\bp) \frac{\bp}{p^0}
\lbl{A3}
\ee
Since $\bp/p^0 = {\bm v}$, the above expression is indeed the velocity operator. Exactly
the same equation (\ref{A3}) is satisfied by the Newton-Wigner position operator\,\ci{WignerPosition}.
In Ref.\,\ci{Teller} it is pointed out that the state
$|\bx \rangle $ created by $a^\dg (\bx)$ as defined in Eq.\,(\ref{3.15a}) is in fact the Newton-Wigner
localized state.

From (\ref{A1}) we find that ${\hat \bx}$ is Hermitian, ${\hat \bx}^\dg = {\hat \bx}$. We will show that
it is self-adjoint with respect to the scalar product in the $\bx$-space.

A single particle wave packet profile is
\be
  |\psi \rangle = \int \dd^3 \bp \, g(\bp) a^\dg (\bp) \vac = \int \dd^3 \bx \, f(\bx) a^\dg (\bx) \vac ,
\lbl{A4}
\ee
where $f(\bx)$ is the Fourier transform of $g(\bp)$.

We define the scalar product according to
\be
  \langle \psi | \psi \rangle = \int \dd^3 \bp \, g^* (\bp) g(\bp) = \int \dd^3 \bx \, f^* (\bx) f(\bx).
\lbl{A5}
\ee
The expectation value of the operator ${\hat \bx}$ is
\be
  \langle \psi |{\hat \bx} |\psi \rangle = \int \dd^3 \bx \, f^* (\bx) \bx f (\bx) .
\lbl{A6}
\ee
We have
\be
   \langle \psi |{\hat \bx} |\psi \rangle^*  =    \langle \psi |{\hat \bx} |\psi \rangle ,
\lbl{A7}
\ee
therefore the operator ${\hat \bx}$ is self-adjoint with respect to the scalar product (\ref{A5}).
Analogous holds for many particle wave packet profiles.

The operators $a^\dg (\bx)$, $a(\bx)$, ${\hat \bx}$ have been defined with respect to
a particular $(1+3)$ split\footnote{
For simplicity reason we take here spatial dimension $\bD = 3$.}
 of spacetime, such that the simultaneity hypersurface $\Sigma_\mu$ coincides with the
space of  coordinates $\bx$, whereas the orthogonal to $\Sigma_\mu$ points into the direction
 of the time coordinates $x^0 \equiv t$. In a different $(1+3)$ split, instead of
 $x^\mu = (t, \bx)$, we have different coordinates $x'^\mu = (t',\bx')$, and thus different
 operators $a^\dg (\bx')$, $a(\bx')$, $\hbx'$. To different $(1+3)$ splits there correspond
 different choices of simultaneity hypersurfaces $\Sigma_\mu$, and thus different
 Lorentz frames. Our operators are thus defined with respect a a given Lorentz frame $S$,
 at a fixed value of time $t$. In the frame $S'$ one has to {\it define} different
 operators, which are suitable creation/annihilation and position operators in $S'$, but
 not in $S$. The theory is thus covariant in the sense
 that we can perform different $(1+3)$ spits and define in each of them the creation/annihilation
 and position operators.  But those operators themselves are not Lorentz covariant objects\footnote{
 It is often said that those quantities are not Lorentz {\it invariant}. But this is misleading, because
 physical quantities need not be Lorentz invariant, they only need be Lorentz {\it covariant};
 from a different Lorentz frame they may look different.}
 and cannot be transformed into another Lorentz reference frame. The transformation of $a^\dg (\bx)$
 defined in Eq.\,(\ref{3.15a}) into another Lorentz frame makes no sense. The same is true
 for the localized state  $|\bx \rangle = a^\dg (\bx) \vac$. It is not correct to say that such a localized
 state in $S$ when observed from $S'$ acquires the strange properties of giving a nonzero amplitude
 for detection spread out all over space. To see how a localized state looks from the frame $S'$,
 one must consider not only the basis state $|\bx \rangle$, but also an amplitude $f(\bx)$.
 In Appendix B we  show that with the aid of the non covariant operators $a^\dg (\bx)$,
 we obtain the appropriate equations for amplitudes, and the corresponding 4-current which is
 a Lorentz covariant object.

 \vs{1cm}
 
 {\bf \large Appendix B: Schr\"odinger equation and the probability current for
 single particle wave packet profiles}
 
 \vs{2mm}
 
 In terms of the creation operators $a^\dg (\bp)$ or $a^\dg (\bx)$, a general single particle
 state can be expressed as
 \be
  |\psi \rangle = \int \dd^\bD \bp \, g(\bp) a^\dg (\bp) \vac = \int \dd^\bD \bx \, f(\bx) a^\dg (\bx) \vac .
\lbl{B1}
\ee
Though we do not explicitly denote so, $g$ and $f$ depend on time $t$ as well.
The Schr\"odinger  equation is
\be
  i \frac{\p |\psi \rangle}{\p t} = H |\psi \rangle
\lbl{B2}
\ee
Taking the Hamiltonian (\ref{A2}), we obtain
\be
    i \frac{\p |\psi \rangle}{\p t} = \int \dd^\bD \bp \,i \,\frac{\p g(\bp)}{\p t} a^\dg (\bp) \vac =
    \int \dd^\bD \bp \,g(\bp) \,\omega_\bp \,a^\dg (\bp) \vac ,
\lbl{B4}
\ee
so that the amplitude $g(\bp)$ satisfies
\be
  i \frac{\p g(\bp)}{\p t} = \omega_\bp \,g(\bp) .
\lbl{B5}
\ee
For the amplitude $f(\bx)$ we have
\bear
   i \frac{\p |\psi \rangle}{\p t} &=& \int \dd^\bD \bx \,i \frac{\p f(\bx)}{\p t} a^\dg (\bx) \vac =
    \int \dd^\bD \bp \,\dd^\bD \bx \,\dd^\bD \bx' f(\bx) \frac{{\rm e}^{i \bp (\bx'-\bx)}}{(2 \pi)^\bD}
    \sqrt{m^2+\bp^2} \, a^\dg (\bx') \vac \nonumber \\
   &=& \int \dd^\bD \bp \,\dd^\bD \bx \,\dd^\bD \bx' f(\bx) \frac{{\rm e}^{i \bp (\bx'-\bx)}}{(2 \pi)^\bD}
   \, m \left (  1 + \frac{\bp^2}{2 m^2} + ...  \right ) a^\dg (\bx') \vac \nonumber \\ 
    &=& \int \dd^\bD \bx \,\dd^\bD \bx' f(\bx) \frac{1}{(2 \pi)^\bD}
   \, m \left (  1 + \frac{(- i { \nabla})^2}{2 m^2} + ...  \right ) \delta^\bD (\bx - \bx') a^\dg (\bx') \vac \nonumber \\
 \lbl{B6}
\ear
After performing a partial integration, the action of the operator
$m \left (  1 + \frac{(- i {\nabla})^2}{2 m^2} + ...  \right )$~ $= (m^2 + (-i {\nabla})^2)^{1/2}$
can be switched from $\delta^\bD (\bx - \bx')$ to $f(\bx)$, so that $\delta^\bD (\bx - \bx')$ becomes
``free", and can be integrated out. So we obtain that the Schr\"odinger equation (\ref{B6}) for a single
particle wave packet profiles is satisfied if the amplitude satisfies\,\ci{RelatSchEquation}--\ci{RelatSchEquation3}
\be
  i \frac{\p f}{\p t} = \left ( m^2 + (-i { \nabla})^2) \right )^{1/2}  f.
\lbl{B7}
\ee
The Hamilton operator in the above equation when expanded contains derivatives up to infinite
order. Therefore the function $f$ satisfying (\ref{B7}), even if initially $f(0,\bx)$
localized within a finite region, at any later time $t$  has non vanishing values
at all points $\bx$. Therefore in the literature \,\ci{RelatSchEquation}--\ci{RelatSchEquation3}
it is usually said that such a wave function is non local.
But in Ref.\,\ci{Al-Hashimi} it was shown that $f(t,\bx)$ which satisfies the initial condition of a minimal uncertainty
in position and momentum evolves as a wave packet whose probability density is concentrated in
a finite spatial region. Using the results of Ref.\,\ci{Al-Hashimi} we have found that sufficiently close to the initial time
the localization of such a wave packet is even more pronounced. The contribution of the wave packet's
tail is small in comparison to the contribution of the region around the wave packet's center.
 
By the way, if $f$ were a spinor, we could take the square root \`a la Dirac, and (\ref{B7})
would become the Dirac equation
\be
  i \frac{\p f}{\p t} = (m \beta + i {\bm \alpha} {\nabla}) f .
\lbl{B8}
\ee
In such a case, of course, $a^\dg (\bx)$ should be replaced by fermionic operators, and instead of the
scalar field theory we would have the fermionic field theory. 

What about the probability density and current? From Eq.\,(\ref{A5}) we see that
\be
  f^* (\bx) f(\bx) = \rho (\bx)
\lbl{B8a}
\ee
can serve as the probability density. Differentiating $\rho(\bx)$ with respect to time, and
using the Schr\"odinger equation (\ref{B7}), we obtain
\be
  {\dot \rho} = i \left (    (H f^*) f - f^* (H f) \right   ) .
\lbl{B9}
\ee
Using the expansion
\be
  H f = \sqrt{m^2 + (-i {\nabla})^2} \, f = m \left ( 1 + \frac{1}{2} \frac{(-i {\nabla})^2}{m^2}
    - \frac{1}{2.4} \frac{(- i {\nabla})^4}{m^4} + ... \right ) f
\lbl{B10}
\ee
we can rewrite Eq.\,(\ref{B9}) in the form
\be
  {\dot \rho} = {\nabla} {\bm j} ,
\lbl{B11}
\ee
where
\be
  {\bm j} = - i m f^* \left [ \frac{1}{2 m^2} ({\stackrel{\leftarrow}{\nabla}} - {\stackrel{\rightarrow}{\nabla}}) 
  + \frac{1}{2.4} \left ( {\stackrel{\leftarrow}{\nabla}}^3 
  - {\stackrel{\leftarrow}{\nabla}}^2 {\stackrel{\rightarrow}{\nabla}}
	   + {\stackrel{\leftarrow}{\nabla}} {\stackrel{\rightarrow}{\nabla}}^2 
	   - {\stackrel{\rightarrow}{\nabla}}^3 \right )  + ... \right ]  f
\lbl{B12}
\ee
Thus we can define the probability density accroding to (\ref{B8a}), but for the corresponding current
we then have the cumbersome expression (\ref{B12}). In the case of the Dirac equation (\ref{B8}),
the current is the usual simple expression entering the 4-current  $j^\mu = (f^\dg f, f^\dg {\bm \alpha} f)$.

For a scalar field, a covariant definition of the probability density can be derived from the
expectation value of the operator ${\hat p}^0 = \left ( m^2 + (-i {\nabla})^2) \right )^{1/2}$:
\bear
  \frac{1}{m} \langle \psi|{\hat p}^0 |\psi \rangle 
  &=& \frac{1}{m} \int \dd^3 \bx f^* (\bx) \sqrt{m^2 + (-i {\nabla})^2}\, f(\bx) 
  = \frac{1}{m} \int \dd^3 \bx \, f^* i \frac{\p}{\p t} f \nonumber \\
  &=& \frac{i}{2 m} \int  \dd^3 \bx \left ( f^*  \frac{\p}{\p t} f - \frac{\p}{\p t} f^* f \right )
  = \int \dd^3 \bx j^0  ,
\lbl{B13}
\ear
where
\be
   j^0 = \frac{i}{2 m}  \left ( f^*  \frac{\p}{\p t} f - \frac{\p}{\p t} f^* f \right ) .
\lbl{B14}
\ee
From the time derivative
\be
  \frac{\p j^0}{\p t} = \frac{i}{2m} (f^* {\ddot f} - {\ddot f}^* f ),
\lbl{B15}
\ee
by using
\bear
   &&i {\dot f} = H f~,~~~~~~- i {\dot f}^* = H f^*~, ~~~~~~~  H = \sqrt{m^2+(-i {\nabla})^2} ,\nonumber\\
   &&{\ddot f} = - H^2 f~, ~~~~~~~{\ddot f}^* = - H^2 f^*~, ~~~~~ H^2 = m^2+(-i {\nabla})^2
\lbl{B16}
\ear
we obtain
\be
   \frac{\p j^0}{\p t} = {\nabla } {\bm j}~,
    ~~~~~{\bm j} = \frac{i}{2m} \left ({\nabla} f^* f - f^* {\nabla} f \right ) .
\lbl{B17}
\ee
We see that the manipulations with position creation/annihilation operators
in the $\bx$-representation are straightforward and lead to the result
that make sense, and are consistent with those in the usual scalar field theory.
The non covariantly defined operators $a^\dg (\bx)$, $a(\bx)$ do not appear
in the definition of the covariant object, the 4-current $j^\mu = (j^0, {\bm j})$, defined
in Eqs. (\ref{B14}), (\ref{B17}).

We will now complete our discussion by considering a state $|\psi \rangle$, defined
according to Eq.\,(\ref{B1}), in which $f(\bx)$ represents a localized state.
Let us consider  the case of a state localized in the initial time $t=0$ at 
position $\bx_0$, so that
\be
  f(\bx) = f(0,\bx) = \delta^\bD (\bx-\bx_0).
\lbl{B18}
\ee
The Fourier transformed wave packet profile is
\be
  g(\bp) = g(0,\bp) = \frac{1}{\sqrt{(2 \pi)^\bD}} \int \dd^\bD \bx \,{\rm e}^{-i \bp \bx} f(\bx) =
  \frac{1}{\sqrt{(2 \pi)^\bD}} {\rm e}^{-i \bp \bx_0} .
\lbl{B19}
\ee
The state at any time is given in terms of $g(t,\bp)$, which evolves according to Eq.\,(\ref{B5}),
the solution being
\be
  g(t,\bp) = {\rm e}^{i \omega_\bp t} g(0,\bp).
\lbl{B20}
\ee
So we have
\be
  |\psi (t) \rangle = \int \dd^\bD \bp \, g(t,\bp) a^\dg (\bp) \vac =
  \int \frac{\dd^\bD \bp}{\sqrt{(2\pi)^\bD}} \, {\rm e}^{i \omega_\bp t} {\rm e}^{-i \bp \bx_0} a^\dg(\bp) \vac .
\lbl{B21}
\ee
The projection of Eq.\,(\ref{B21}) into the state $\langle \bp |= \langle 0| a(\bp)$ gives
\be
  \langle \bp |\psi(t) \rangle = \frac{1}{\sqrt{(2 \pi)^\bD}} \, {\rm e}^{i \omega_\bp t} {\rm e}^{-i\bp \bx_0}
  = g(t,\bp) ,
\ee
whereas the projection into the state $\langle \bx| = \langle 0| a(\bx)$ gives
\be
  \langle \bx|\psi(t) \rangle = f(t,\bx) = \frac{1}{(2 \pi)^\bD} \int \dd^\bD \bp \,
  {\rm e}^{i \sqrt{m^2 + \bp^2} t} . {\rm e}^{i (\bp (\bx - \bx_0)} =
  {\rm e}^{i\sqrt{m^2 + (-i \nabla)^2} t} \delta^\bD (\bx-\bx_0).
\lbl{B23}
\ee
Expanding $\sqrt{m^2 + \bp^2} = m \left ( 1 + \frac{bp^2}{2 m^2} + ... \right )$ and neglecting higher
order terms, the above equation gives
\be
  f(t,\bx) = \frac{1}{(2 \pi)^\bD} \int \dd^\bD \bp \, {\rm e}^{i m t} 
  {\rm e}^{i \frac{\bp^2}{2 m} t}  {\rm e}^{i (\bp (\bx - \bx_0)}  ,
\lbl{B24}
\ee
which, apart form the phase factor ${\rm exp} (i m t)$, is just the Green function for
a non relativistic free particle. The full expansion gives (\ref{B23}), which is the Green function
for a single free relativistic particle\,\ci{Al-Hashimi}.

Instead of the $\delta$-function initial localization (\ref{B18}), we can take a Gaussian function
\be
  f(0,\bx) \propto {\rm exp} \left [ \frac{(\bx-\bx_0)^2}{2 \sg_0} \right ] .
\lbl{B23a}
\ee
This holds in $S$. Observed from another reference frame $S'$, moving with respect to $S$
with velocity $v$ along the $x^1 \equiv x$ direction, the wave function  transform according to\footnote{
We now take the 4D spacetime, and denote $x^1 \equiv x$, $x^2 \equiv y$, $x^3 \equiv z$.}
\be
  f(t,\bx) = f'(t',\bx') \propto {\rm exp} \left [ (1-v^2)(x'-x_0')^2 +
  (y'-y_0')^2 + (z'-z_0')^2 \right ] ,
\lbl{B24a}
\ee
where $t=0$ and $t' = - \frac{v x'}{\sqrt{1-v^2}}$. This is also a localized wave packet,
only the distance is Lorentz contracted and different points $x'$ are not simultaneous
in $S'$. In deriving (\ref{B24a}) we used  the transformations
\bear
  &&x = \frac{x' + v t'}{\sqrt{1-v^2}}~,~~~~~~~~~x_0 =  \frac{x_0' + v t_0'}{\sqrt{1-v^2}}~,
  ~~y=y'~,~~z=z' \lbl{B25}\\
  &&t = \frac{t' + v x'}{\sqrt{1-v^2}}~,~~~~~~~~~t_0 =  \frac{t_0' + v x_0'}{\sqrt{1-v^2}}
  \lbl{B26}
\ear
Taking $t=t_0$ (which means that both points, $x$ and $x_0$, are simultaneous in $S$),
Eq.\,(\ref{B26}) gives $t' - t_0' = - v (x' - x_0')$ (which means that those two points are
not simultaneous in $S'$). From Eq.\,(\ref{B25}) we then obtain
$x-x_0 = \sqrt{1-v^2}(x' - x_0')$, i.e., the length contraction along the $x$-direction.

We see that nothing strange happens with the particle localization if we observe
it from another Lorentz frame. Instead of the spherical Gaussian (\ref{B23a}) at
$t=0$, we see in $S'$ an ellipsoidal Gaussian (\ref{B24a}) function, each $x'$ taken
at different $t'= - v x'/\sqrt{1-v^2}$. This reflects the fact that in $S$ the localization
is on the simultaneity hyper surface $t=0$, which in $S'$ is not a simultaneity
Hyper surface. The observer in $S'$, of course, normally does not define the spread
of a localized particle's position
at different values of his time $t'$, he defines it at the same value of $t'$,
i.e., on the simultaneity hyper surface of  the Lorentz frame $S'$. Therefore the
observer in $S'$ formulates the quantum field theory with respect to $S'$ in the
same manner as we did with respect to $S$. In $S'$ one then obtains results concerning
wave packets and localization that are analogous to those that we obtained
in the reference frame $S$. As already mentioned, according to Ref.\,\ci{Al-Hashimi}
such an initially Gaussian wave packet evolves so that it remains localized in the
sense that the contribution of its tail remains small in comparison to the
contribution around the wave packet's center. Though the tail contains superluminal
propagation, it does not necessarily mean the violation of causality, if the latter
is properly defined in terms of macroscopic modulated beams that can bear
information. Single particle events at space-like separations can not transmit
information, and therefore do not violate the properly defined causality.

 \vs{1cm}
 
 {\bf \large Appendix C: Comparison with the Newt-Wigner position operator}
 
 \vs{2mm}
 
If instead of the operators $a(\bp)$, $a^\dg (\bp)$ satisfying the commutation
relations
\be
  [a(\bp),a^\dg(\bp)] = \delta^3 (\bp- \bp') ,
\lbl{C1}
\ee
we use the operator ${\tl a} (\bp) = \sqrt{(2 \pi)^3 2 \omega_\bp }\, a(\bp)$,
${\tl a}^\dg (\bp) = \sqrt{(2 \pi)^3 2 \omega_\bp }\, a^\dg (\bp)$ satisfying
\be
  [{\tl a}(\bp),{\tl a}^\dg(\bp)] = (2 \pi)^3 2 \omega_\bp \,\delta^3 (\bp- \bp') ,
\lbl{C2}
\ee 
then the position operator (\ref{A1}) becomes
\be
  \hbx = \int \frac{\dd^3 \bp} {(2 \pi)^3 2 \omega_\bp} \, {\tl a}^\dg (\bp)
  i \left ( \nabla_\bp - \frac{\bp}{2 \omega_\bp^2} \right ) {\tl a} (\bp),
\lbl{C3}
\ee
where we now use $\nabla_\bp \equiv \p/\p \bp$.

Let us consider the state
\be
  |\psi \rangle = \int \frac{\dd^3 \bp} {(2 \pi)^3 2 \omega_\bp} {\tl g}(\bp) {\tl a}^\dg (\bp) \vac ,
\lbl{C4}
\ee
act on it by the operator $\hbx$, and project $\hbx |\psi \rangle$ onto the state ${|{\tl \bx} \rangle}$
defined as
\be
  |{\tl \bx} \rangle = \varphi^+ (0,\bx) \vac = \int \frac{\dd^3 \bk}{(2 \pi)^3 2 \omega_\bk}\,
  {\rm e}^{- i \bk \bx} {\tl a}^\dg (\bk) \vac ,
\lbl{C5}
\ee
where $\vphi^+ (0,\bx)$ is the positive energy Klein-Gordon field operator. We obtain
\bear
  \langle {\tl \bx}| \hbx |\psi \rangle &=& \int \frac{\dd^3 \bp}
  {(2 \pi)^3 2 \omega_\bp} \,  {\rm e}^{- i \bp \bx} 
  \, i \left ( \nabla_\bp - \frac{\bp}{2 \omega_\bp^2} \right ) {\tl g} (\bp) \nonumber \\
  &=& \left ( \bx + \frac{1}{2 (m^2 + (-i \nabla)^2)} \nabla \right ) {\tl f} (\bx) ,
\lbl{C7}
\ear
where
\be
  {\tl f} = \int \frac{\dd^3 \bp}{(2 \pi)^3 2 \omega_{\bp}}\, {\rm e}^{- i \bp \bx}\, {\tl g} (\bp) .
\lbl{C8}
\ee
In Eq.\,(\ref{C7}) we have precisely the well-known non local action of Newton-Wigner
position operator on a wave function ${\tl f} (\bx)$.

Why then in our paper we do not have such a non local action of the position
operator? Let us rewrite the state (\ref{C4}) in terms of the operators
$a^\dg (\bp) = {\tl a} (\bp)/\sqrt{(2 \pi)^3 2 \omega_\bp}$. If we introduce
$g(\bp) = {\tl g} (\bp)/\sqrt{(2 \pi)^3 2 \omega_\bp}$, then (\ref{C4}) becomes
\be
  |\psi \rangle = \int \dd^3 \bp \, g(\bp) \, a^\dg (\bp) \vac ,
\lbl{C9}
\ee
which is just our wave packet (\ref{B1}). But the state $|{\tl \bx} \rangle
 = \vphi^\dagger (0,\bx) \vac$ defined according to Eq.\,(\ref{C5}) is not the same
 as the position state $|\bx \rangle = a^\dg(\bx) \vac$ used throughout
 this paper. If we project $\hbx |\psi \rangle$ onto $|\bx \rangle$, then
\be
 \langle \bx |\hbx |\psi \rangle = \int \frac{\dd^3 \bp \, 
 {\rm e}^{i \bp \bx}}{\sqrt{(2 \pi)^3}}\, i \nabla_\bp g(\bp) = \bx f(\bx) ,
\lbl{C10}
\ee
where
\be
  f(\bx) = \frac{1}{\sqrt{(2 \pi)^3}} \int \dd^3 \bp \,{\rm e}^{i \bp \bx} g(\bp) .
\lbl{C11}
\ee

It is well known that the states $|{\tl \bx} \rangle $ created by the Klein-Gordon
field operator  are not position states. If one uses such states, then
$\langle {\tl \bx}|\hbx |\psi \rangle$ of course gives a non local result,
which is indeed confirmed in Eq.\,(\ref{C7}). If, on the contrary, one uses
the states $|\bx \rangle$ created by $a^\dg (\bx)$, then 
$\langle  \bx |\hbx |\psi \rangle$ gives the position eigenstates,
as shown in Eq.\,(\ref{C10}). The states $|\bx \rangle = a^\dg (\bx) \vac$
are the basis states, defined with respect to a given Lorentz reference
frame. They are not dynamical objects of the theory, they do not satisfy
the Schr\"odinger equation. On the other hand, the state $|\psi \rangle$,
whose definition (\ref{C9}) is equivalent to the definition (\ref{C4}),
must satisfy the Schr\"odinger equation. The state $|\psi \rangle$ is
a Lorentz covariant object. It is the state $|\psi \rangle$, not $|\bx \rangle$,
which is physically relevant, and whose properties (e.g., localization,
behaviour in different Lorentz frames, etc.)  are to be considered.

\vs{7mm}

\centerline{\bf Acknowledgement}

This work has been supported by the Slovenian Research Agency.

\end{document}